\RequirePackage{fix-cm}
\documentclass[smallextended]{svjour3}       
\smartqed  
\usepackage{graphicx}
\usepackage{mathrsfs}
\usepackage{subfigure}
\usepackage{rotating}
\usepackage{color}
\usepackage{amsmath,amssymb}
\usepackage{braket}
\begin{document}

\title{Multiphase lattice Boltzmann simulations for porous media applications}
\subtitle{A review}


\author{
Haihu~Liu
\and
Qinjun~Kang
\and
Christopher~R.~Leonardi
\and
Sebastian~Schmieschek
\and
Ariel~Narv\'aez
\and
Bruce~D.~Jones
\and
John~R.~Williams
\and
Albert~J.~Valocchi
\and
Jens~Harting
}


\institute{
Haihu Liu \at
School of Energy and Power Engineering, Xi'an Jiaotong University, 28 West Xianning Road, Xi'an 710049, China,
\email{haihu.liu@mail.xjtu.edu.cn}
\and
Qinjun Kang\at
Earth and Environmental Sciences Division, Los Alamos National Laboratory, Los Alamos, New Mexico 87545, USA,
\email{qkang@lanl.gov}
\and
Christopher R. Leonardi \at
The University of Queensland,
School of Mechanical and Mining Engineering,
Cooper Road, St Lucia QLD 4072, Australia,
\email{c.leonardi@uq.edu.au}\\
Massachusetts Institute of Technology,
Department of Civil and Environmental Engineering,
77 Massachusetts Avenue, Cambridge MA 02139, USA
\and
Sebastian Schmieschek\at
Centre for Computational Science, Department of Chemistry, University College
London, WC1H 0AJ London, UK\\
Department of Applied Physics, Eindhoven University of Technology, Den Dolech 2, \\5600MB Eindhoven, The Netherlands,
\email{s.schmieschek@ucl.ac.uk}
\and
Ariel Narv\'aez\at
Department of Applied Physics, Eindhoven University of Technology, Den Dolech 2, \\5600MB Eindhoven, The Netherlands,
\email{ariel.narvaez@gmx.de}
\and
Bruce D. Jones and John R. Williams \at
Massachusetts Institute of Technology,
Department of Civil and Environmental Engineering,
77 Massachusetts Avenue, Cambridge MA 02139, USA,
\email{bdjones@mit.edu, jrw@mit.edu}
\and
Albert J. Valocchi \at
Department of Civil and Environmental Engineering,
University of Illinois at Urbana-Champaign,
205 N. Mathews Ave.,
Urbana IL 61801, USA\\
International Institute for Carbon Neutral Energy Research (WPI-I2CNER),
Kyushu University, 744 Moto-oka, Nishi-ku, Fukuoka 819-0395, Japan,
\email{valocchi@illinois.edu}
\and
Jens Harting\at
Research Centre Juelich GmbH, Helmholtz-Institute Erlangen-Nuremberg (IEK-11), Fuerther Strasse 248, 90429 Nuremberg, Germany\\
Department of Applied Physics, Eindhoven University of Technology, Den Dolech 2, \\5600MB Eindhoven, The Netherlands,
\email{j.harting@fz-juelich.de}
}

\date{Received: date / Accepted: date}

\maketitle

\begin{abstract}
Over the last two decades, lattice Boltzmann methods have become an
increasingly popular tool to compute the flow in complex geometries such as
porous media. In addition to single phase simulations allowing, for example, a
precise quantification of the permeability of a porous sample, a number of
extensions to the lattice Boltzmann method are available which allow to study
multiphase and multicomponent flows on a pore scale level. In this article we
give an extensive overview on a number of these diffuse interface models and
discuss their advantages and disadvantages. Furthermore, we shortly report on
multiphase flows containing solid particles, as well as implementation details
and optimization issues.
\keywords{Porous media \and Pore scale simulation \and Lattice Boltzmann method}
\PACS{
47.11.-j \and     
91.60.Np \and    
47.56.+r     
}
\end{abstract}

\section{Introduction}
\label{intro}
Fluid flow in porous media is a topic which is relevant in the context of
hydrocarbon production, groundwater flow, catalysis or the gas diffusion layers
in fuel cells~\cite{Hil96}. Oil and gas transport in porous
rock~\cite{PhysRevB.45.7115}, the flow in underground reservoirs and the
propagation of chemical contaminants in the vadose
zone~\cite{2005EG:DN,citeulike:481254}, permeation of ink in
paper~\cite{koponen:3319} and filtration and sedimentation
operations~\cite{2003IJMP:GCB} are just a few examples from a wealth of
possible applications. Most of these examples involve not only single phase
flows, but multiple phases or fluid components.  As such, a thorough
understanding of the underlying physical processes by means of computer
simulations requires accurate and reliable numerical tools.

Multiphase flows in porous media are typically modeled using macro-scale
simulations, in which the continuity equation together with momentum and
species balances are solved and constitutive equations such as Darcy's law are
utilized. These models are based on the validity of the constitutive
relationships (e.g. the multiphase extension of Darcy's Law), require some
inputs for semi-empirical parameters (e.g. relative permeability), and have
difficulties in accounting for heterogeneity, and complex pore interconnectivity
and morphologies~\cite{Balhoff2008}. As a result, macroscale simulations do not
always capture effects associated with the microscale structure in multiphase
flows.
On the contrary, pore-scale simulations are able to capture
heterogeneity, interconnectivity, and non-uniform flow behavior (e.g. various
fingerings) that cannot be well resolved at the macroscopic scale. In
addition, pore-scale simulations can provide detailed local information on
fluid distribution and velocity, and enable the construction and
testing of new models or constitutive equations for macroscopic scales.

Pore-network
models~\cite{Blunt1991,Bryant1992,Al-Gharbi2005,Valvatne2005,Piri2005,Blunt2013,Joekar:2008,Joekar:2010,Raoof:2010}
are a viable tool for understanding multiphase flows at the pore scale, and
they are computationally efficient. These models, however, are based upon
simplified representations of the complex pore geometry~\cite{Jiang:2007}, which
restricts their predictive capability and accuracy.

Traditional CFD methods
such as the volume-of-fluid (VOF)
method~\cite{Hirt1981,Rider1998,Gueyffier1999,Ferrari2013} and level set (LS)
method~\cite{Osher1988,Sussman1998,Osher2003} simulate multiphase flows by
solving the macroscopic {Navier-Stokes equations} together with a proper
technique to track/capture the phase interface. It is challenging to use VOF
and LS methods for pore-scale simulations of multiphase flows in porous media
because of the difficulties in modeling and tracking the dynamic phase
interfaces. Also, they have difficulties incorporating fluid-solid interfacial
effects (e.g. surface wettability) in complex pore structures, which are
consequences of microscopic fluid-solid interactions. 

Unlike traditional CFD methods, which are based on the solution of macroscopic
variables such as velocity, pressure, and density, the lattice Boltzmann method (LBM) is a
pseudo-molecular method that tracks the evolution of the particle distribution
function of an assembly of molecules and is built upon microscopic models and
mesoscopic kinetic equations~\cite{bib:benzi-succi-vergassola,bib:qian-dhumieres-lallemand,bib:succi-01}.
The macroscopic variables are obtained from moment integration of the particle
distribution function.  Even shortly after its introduction more than twenty
years ago, the LBM became an attractive alternative
to direct numerical solution of the Stokes equation for single-phase flows in porous media
and complex geometries in
general~\cite{1990PhFl.2.2085C,bib:qian-dhumieres-lallemand,bib:ferreol-rothman}.
In the LBM for multiphase flow simulations, the fluid-fluid
interface is not a sharp material line, but a diffuse interface of finite width.
The effective slip of the contact line is caused by the relative diffusion of the
two fluid components in the vicinity of the contact line. Therefore, there are no singularities
in the stress tensor in the lattice Boltzmann simulation of moving contact-line problems while the no-slip
condition is satisfied~\cite{Briant2004,Briant2004a,Chen2000b,Kang2002,Kang2004,Kang2005}.
In addition, unlike traditional CFD methods, there is no need for complex
interface tracking/capturing/resconstruction techniques in the diffuse interface methods. Rather, the formation, deformation,
and transport of the interface emerge through the simulation results~\cite{Chen1998}. Furthermore,
in the LBM all computations involve only local variables
enabling highly efficient parallel implementations based on simple domain
decomposition~\cite{bib:jens-harvey-chin-venturoli-coveney:2005}. With more
powerful computers becoming available it was possible to perform detailed
simulations of flow in artificially generated
geometries~\cite{koponen:3319,bib:jens-zauner-weeber-hilfer:2008,bib:jens-narvaez:2010,bib:jens-narvaez-zauner-raischel-hilfer:2010},
tomographic reconstructions of sandstone
samples~\cite{bib:ferreol-rothman,Martys99largescale,MAKHT02,bib:jens-venturoli-coveney:2004,Ahren06},
or fibrous sheets of paper~\cite{koponen:716}.

The remainder of this article is organised as follows: after a more detailed
introduction to the LBM in
Sec.~\ref{sec:LB}, we review a number of different diffuse interface multiphase
and multicomponent models in Sec.~\ref{sec:mult}. Sec.~\ref{sec:mult} also
introduces how particle suspensions can be simulated using the LBM.
Sec.~\ref{sec:imp} summarizes a few typical details to be taken care of when
implementing a lattice Boltzmann code and Sec.~\ref{sec:app} is comprised of a
collection of possible applications of the several multiphase/multicomponent
models available. Sec.~\ref{sec:summ} summarizes our findings and the
advantages and limitations of the various methods.

\section{The lattice Boltzmann method}\label{sec:LB}
The LBM can be seen as the successor of the lattice
gas cellular automaton (LGCA) which was first proposed in 1986 by Frisch,
Hasslacher, and Pomeau~\cite{bib:fhp}, as well as by
Wolfram~\cite{bib:wolfram}. The LBM overcomes some limitations of the LGCA such
as not being Galilei-invariant and numerical noise due to the Boolean nature of the algorithm.  In contrast to the LGCA coarse
graining of the molecular processes is not obtained by tracking individual
discrete mesoscopic fluid packets anymore. Instead, in the LBM the dynamics of
the single-particle distribution function $f(\vec{x},\vec{v},t)$ representing
the probability to find a fluid particle with position $\vec{x}$ and velocity
$\vec{v}$ at time $t$ is
tracked~\cite{mcnamara_use_1988,higuera_boltzmann_1989,benzi_lattice_1992,chen_recovery_1992,qian_lattice_1992}.
Then, the density and velocity of the macroscopically observable fluid are
given by $\rho(\vec{x},t) = \int f {\mathrm d}\vec{v} $ and $\vec{u}(\vec{x},t)
= \int f \vec{v} {\mathrm d} \vec{v}$, respectively. In the non-interacting,
long mean free path limit and with no externally applied forces, the evolution of
this function is described by the Boltzmann equation,
\begin{equation}
\label{eq:boltzmann}
\left( \partial_t + \vec{v} \cdot \vec{\nabla} \right) f
= \Omega[f].
\end{equation}
The left hand side describes changes in the distribution function due to
free particle motion. The collision operator $\Omega$ on the right hand side
describes changes due to pairwise collisions.
In general, this is a complicated integral expression, but it is commonly
simplified to the linear Bhatnagar-Gross-Krook, or BGK
form~\cite{bib:bgk},
\begin{equation}
\label{eq:bgk}
\Omega[f] \simeq - \frac 1 \tau \left[ f - f^{\mathrm{(eq)}} \right].
\end{equation}
This collision operator describes the relaxation towards a Maxwell-Boltzmann
equilibrium distribution $f^{\mathrm{(eq)}}$ at a time scale set by the
characteristic relaxation time $\tau$. The
distributions governed by the Boltzmann-BGK equation conserve
mass, momentum, and energy, and obey a non-equilibrium form of the second law
of thermodynamics~\cite{bib:liboff}. Moreover, the Navier-Stokes equations for
macroscopic fluid flow are obeyed in the limit of small Knudsen and Mach
numbers (see below)~\cite{bib:chapman-cowling,bib:liboff}.

By discretizing the single-particle distribution in time and space, the lattice
Boltzmann formulation is obtained. Here, the positions $\vec{x}$ on which $f$ is
defined are restricted to nodes of a lattice, and the velocities
are restricted to a set $\vec{e}_i, i=1,...,N$ joining these nodes.
$N$ varies betweeen implementations and we refer
to the article of Qian~\cite{qian_lattice_1992} for an overview. We restrict
ourselves to the popular D2Q9 and D3Q19 realizations, which correspond to a 2D
lattice with 9 possible velocities and a 3D lattice with 19
possible velocities, respectively.
To simplify the notation, $f_i(\vec{x},t) = f(\vec{x},\vec{e}_i,t)$ represents the probability to find
particles at a lattice site $\vec{x}$ moving with velocity $\vec{e}_i$, at
the discrete timestep $t$. The density and momentum of the simulated fluid are calculated as
\begin{equation}
\label{eq:lbe-density}
\rho(\vec{x},t) = \rho_0\sum_i f_i(\vec{x},t),
\end{equation}
and
\begin{equation}
\label{eq:lbe-velocity}
\rho(\vec{x},t)\vec{u}(\vec{x},t) = \sum_i f_i(\vec{x},t) \vec{e}_i,
\end{equation}
where $\rho_0$ refers to a reference density which is kept at $\rho_0=1$ in the
remainder of this article. The pressure of the fluid is calculated via an isothermal
equation of state,
\begin{equation}
\label{eq:lbe-pressure}
\ p = c_s^{2}\rho.
\end{equation}
Here, $c_s = c/\sqrt{3}$ is the lattice speed of sound and $c =
\delta_x/\delta_t$ is the lattice speed. The lattice must be chosen carefully
to ensure isotropic behavior of the simulated
fluid~\cite{bib:qian-dhumieres-lallemand}. The lattice Boltzmann formulation
can be obtained using alternative routes, including discretizing the continuum
Boltzmann equation~\cite{bib:he-luo}, or regarding it as a Boltzmann-level
approximation of the LGCA~\cite{bib:mcnamara-zanetti}.

The LBM follows a two-step procedure, namely an
advection step followed by a collision step. In the advection step, values of
the distribution function are propagated to adjacent lattice sites along their
velocity vectors. This corresponds to the left-hand side of the continuum
Boltzmann equation. In the collision step, particles at each lattice site are
redistributed across the velocity vectors. This process corresponds to the
action of the collision operator, and in the most simple case takes the BGK
form. The combination of the advection and collision steps results in the lattice Boltzmann equation (LBE),
\begin{equation}\label{eq:LBE}
f_i(\vec{x}+\vec{e}_i\delta_t,t+\delta_t)-f_i(\vec{x},t) = \Omega_i(\vec{x},t).
\end{equation}
In most applications and the remainder of this article the reference density, timestep and lattice constant are chosen to be $\rho_0=1$, $\delta_t=1$ and $\delta_x=1$.
The discretized local equilibrium distribution is often given by a second order Taylor expansion of the Maxwell-Boltzmann equilibrium distribution,
\begin{equation}
f_i^{eq} =  w_i \rho \left(1+\frac{1}{c_{s}^2}\vec{e}_i\cdot \vec{u}+\frac{1}{2c_{s}^4}(\vec{e}_i\cdot
    \vec{u})^2-\frac{1}{2c_{s}^2}|\vec{u}|^2\right).
\end{equation}
Therein, the coefficients including the weights $w_i$ associated to the lattice
discretisation and the speed of sound $c_{s}$ are determined by a comparison of
a first order Chapman-Enskog expansion to the Navier-Stokes equations. The
kinematic viscosity of the fluid,
\begin{equation}\label{eq:nu}
\nu = {c_s^2}\left(\tau-\frac{1}{2}\right),
\end{equation}
is determined by the relaxation parameter $\tau$. 

While the simplicity of the LBGK method has enabled it to be successfully
applied to a wide range of problems
\cite{chen_lattice_1998,bib:succi-01,aidun_lattice-boltzmann_2010}, it also
implies limitations to the formalism. The implicit relationship between fluid
properties and discretization parameters in Eq.~\ref{eq:nu} leads to numerical
instability at lower viscosities~\cite{lallemand_theory_2000}. As indicated by
the equation of state in Eq.~\ref{eq:lbe-pressure}, the LBM approximates the
Navier-Stokes equations in the near-incompressible limit. To minimise
compressibility errors, and to adhere to the small-velocity assumption, the
Mach number, $Ma = u/c_s$, has to be kept small (i.e. $Ma \ll 1$).

To address some of these limitations, different approaches and extensions to
the formalism have been introduced. At an early stage of the LBM's development,
alternative collision schemes were
introduced~\cite{bib:higuera-succi-benzi,dhumieres_generalized_1994}. In
particular, the multiple relaxation time (MRT) collision operator can be
written
as~\cite{ladd_numerical_1994,dhumieres_generalized_1994,dhumieres_multiplerelaxationtime_2002},
\begin{equation}
  \Omega_i^{MRT}(\vec{x},t) = - \mathcal{M}^{-1} \mathcal{\hat{S}} \mathcal{M} \left[\ket{f(\vec{x},t)} - \ket{f^{eq}(\vec{x},t)}\right].
\end{equation}
Herein $\mathcal{M}$ is an invertible transformation matrix, relating the
moments of the single particle velocity distribution $f$ to linear combinations
of its discrete components $f_{i}$. It can be obtained by a Gram Schmidt
orthogonalization of a matrix representation of the stochastical moments.  The
collision process is performed in the space of moments, where
$\mathcal{\hat{S}}$ is a diagonal matrix of the individual relaxation times.
Thus, independent transport coefficients are introduced. For example, in
addition to the shear viscosity, the bulk viscosity
$\xi=c_s^2\left({\tau_{\mathsf{bulk}}}-\frac{1}{2}\right)$ can be
controlled~\cite{dhumieres_multiplerelaxationtime_2002}.

Starting from this general approach, simplifications and extensions have led to
the development of, for example, two relaxation time (TRT)
models~\cite{ginzburg_equilibrium-type_2005,ginzburg_optimal_2010} as well as
models incorporating thermal
fluctuations~\cite{ladd_short-time_1993,dunweg_statistical_2007,gross_thermal_2010,kaehler_fluctuating_2013}.


Further refinement of the method has been achieved by identifying general
formalisms for deriving higher order expansions of the equilibrium distribution
and lattice discretisations allowing to include higher order effects into the
model~\cite{shan_kinetic_2006,dubois_towards_2009}.

The ease in handling boundaries is one of the reasons for the LBM being well
suited to simulating porous media flows. Many boundary condition
implementations maintain the locality of LBE operations, which means that
tortuous pore network geometries can be modeled on an underlying orthogonal
grid, and that parallelization of the method remains straightforward.

\begin{figure}
  \centering
  \subfigure[]
  {\label{fig:microCTrender}\includegraphics[width=0.45\textwidth]{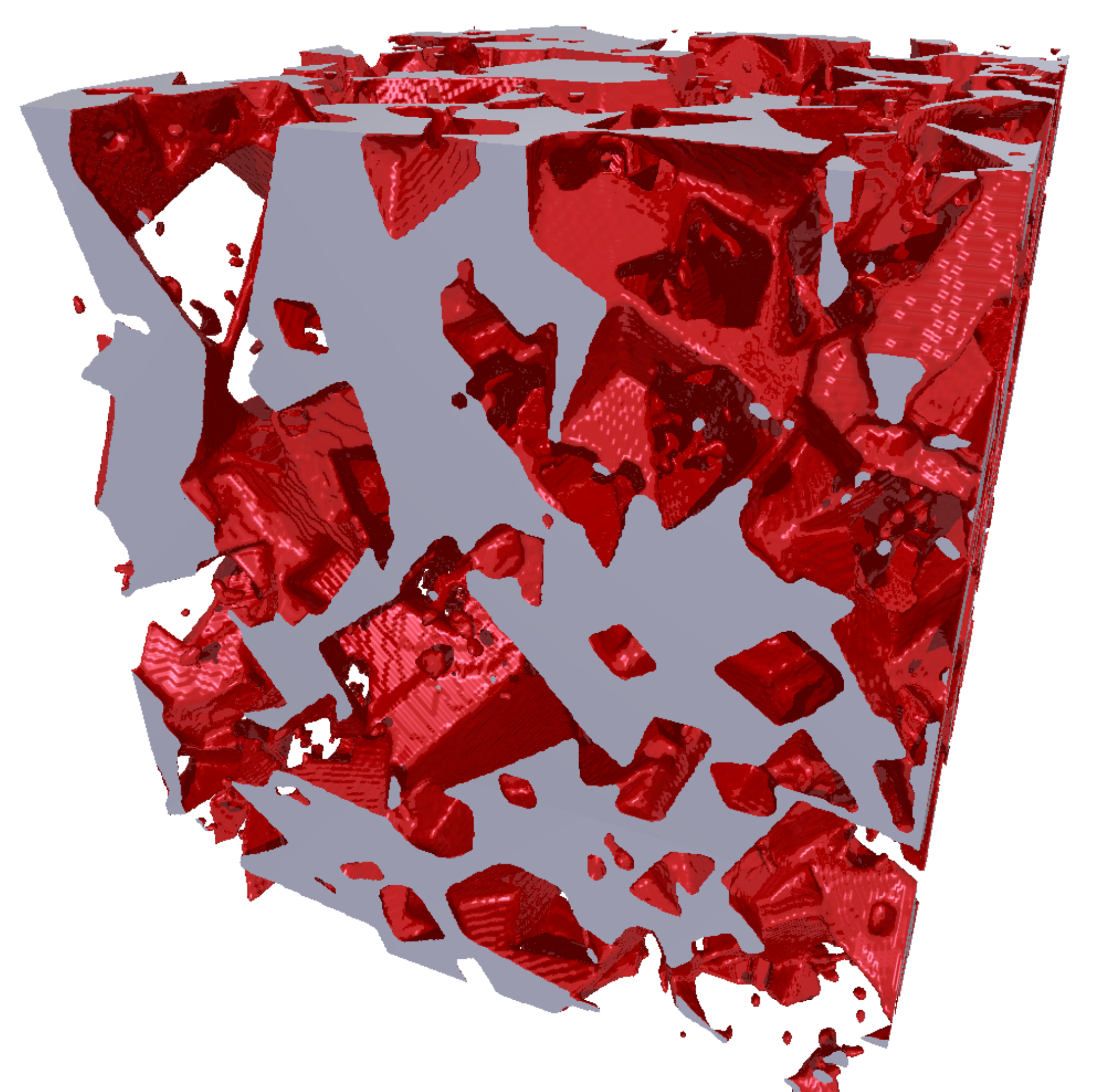}}
  \hfill
  \subfigure[]
  {\label{fig:microCTflow}\includegraphics[width=0.45\textwidth]{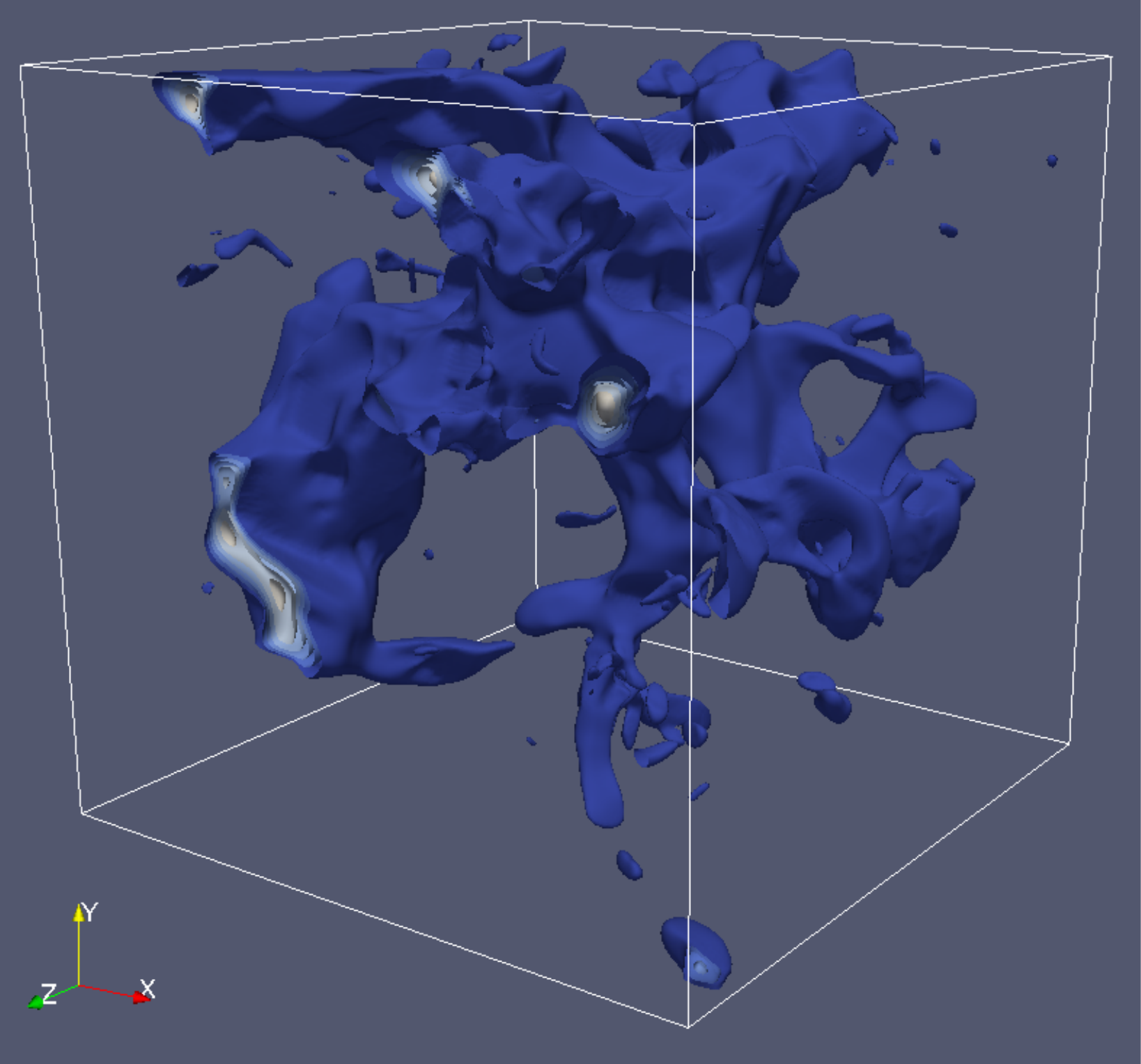}}
  \caption{Single phase flow in a segmented, $\mu$CT image of a Dolomite sample including \subref{fig:microCTrender} a rendering of the pore volume (i.e. the complement of the rock volume) in the sample and \subref{fig:speedup-md2} the steady state flow profile in the sample as computed by the LBM with bounce-back boundary conditions.}
\end{figure}

The simplest approach to model the interaction of fluid and solid is the
bounce-back scheme.
It enforces the no-slip condition at solid surfaces by
reflecting particle distribution functions from the boundary nodes back in the
direction of incidence. Advantages of the bounce-back condition are that
the required operations are local to a node and that the orientation of the
boundary with respect to the grid is irrelevant. However, the simplicity of the
bounce-back scheme is at the expense of accuracy. It has been shown that
generally it is only first-order in numerical accuracy~\cite{Cornubert1991} as
opposed to the second order accuracy of the lattice Boltzmann equation at
internal fluid nodes~\cite{bib:chen-doolen}. It has also been
shown~\cite{bib:pf.QZoXHe.1997} that the bounce-back condition actually results
in a boundary with a finite relaxation time dependent slip~\cite{narvaez_quantitative_2010}.
Nevertheless, the bounce-back scheme is usually suitable
for simulating the fluid interaction at stationary boundaries such as the
Dolomite rock sample shown in Fig.~\ref{fig:microCTrender}.

Pressure and velocity boundary conditions can be applied in the LBM by
assigning particle distribution functions at a node which correspond to the
prescribed macroscopic constraint. As an example, Zou and
He~\cite{bib:pf.QZoXHe.1997} proposed a boundary condition based on
bouncing-back the non-equilibrium part of the distribution function. It can be
applied to velocity, pressure and wall constraints. As with the bounce-back
condition, all required operations are local. While the original implementation
was limited to two dimensions and boundaries parallel to the
orthogonal lattice directions, Hecht and Harting presented how to overcome
these limitations~\cite{HH08b}. Periodic and stress-free boundary
implementations are also available, and a detailed review of other velocity
boundary condition implementations in the LBM can be found in~\cite{Latt2008}.

\section{Review of Multiphase$/$Multicomponent LBM Formulations }
\label{sec:mult}
A number of multiphase LBM models have been proposed in the literature. Among them,
five representative models are the color gradient
model~\cite{Gunstensen1991,Reis2007,Liu2012a}, the inter-particle potential
model~\cite{Shan1993,Shan1994,Shan1995}, the free-energy
model~\cite{Swift1995,Swift1996}, the mean-field theory model~\cite{He1999},
and the stabilized diffuse-interface model~\cite{Lee2010}.  In this section, we
review these models with emphasis on some recent improvements and show
their advantages and limitations for pore-sale simulation of multiphase flows in porous
media.

\subsection{The color gradient model}
The color gradient model originated from the two-component lattice-gas model proposed by
Rothman \& Keller~\cite{Rothman1988}, and was first introduced by Gunstensen et
al.~\cite{Gunstensen1991} for simulating immiscible binary fluids based on a
two-dimensional(2D) hexagonal lattice. Later, it was modified by Grunau et
al.~\cite{Grunau1993} to allow variations of density and viscosity. In this
model, ``Red" and ``Blue" distribution functions $f_i^R$ and $f_i^B$ were
introduced to represent two different fluids. The total particle distribution
function is defined as: $f_i=f_i^R+f_i^B$. Each of the colored fluids undergoes
the collision and streaming steps,
\begin{eqnarray}
f_i^{k \dag}(\vec{x},t)=f_i^k(\vec{x},t)+\Omega_i^k(\vec{x},t), \\
f_i^k(\vec{x}+\vec{e}_i\delta_t,t+\delta_t)=f_i^{k \dag}(\vec{x},t),
\end{eqnarray}
where the superscript $k=R$ or $B$ denotes the color (``Red" or
``Blue"), and the collision operator $\Omega_i^k$ consists of three sub-operators~\cite{Tolke2002a,Liu2012a},
\begin{equation}
\label{eq_colli}
    \Omega_i^k=(\Omega_i^k)^{(3)}\left[(\Omega_i^k)^{(1)}+(\Omega_i^k)^{(2)}\right].
\end{equation}

In Eq.(\ref{eq_colli}), $(\Omega_i^k)^{(1)}$ is the BGK collision operator, defined as $(\Omega_i^k)^{(1)}=-\frac{1}{\tau_k}(f_i^k - f_i^{k,eq})$, where $\tau_k$ is the dimensionless relaxation time of fluid $k$, and $f_i^{k,eq}$ is the equilibrium distribution function of
$f_i^k$. Conservation of mass for each fluid and total momentum
conservation require,
\begin{equation}
\label{eq_rhok}
   \rho_k=\sum_i f_i^k=\sum_i f_i^{k,eq},
\end{equation}

\begin{equation}
\label{eq_uk}
   \rho\vec{u}=\sum_i\sum_k f_i^k\vec{e}_i=\sum_i\sum_k
   f_i^{k,eq}\vec{e}_i,
\end{equation}
where $\rho_k$ is the density of fluid $k$, $\rho=\rho_R+\rho_B$
is the total density and $\vec{u}$ the local fluid velocity.

$(\Omega_i^k)^{(2)}$ is a two-phase collision
operator (i.e. perturbation step) which contributes to the mixed
interfacial region and generates an interfacial tension. For a 2D hexagonal lattice, the perturbation operator is given as~\cite{Gunstensen1991,Grunau1993},
\begin{equation}\label{eq_d2q7}
(\Omega_i^k)^{(2)}=\frac{A_k}{2}|\vec{G}|\left[\frac{(\vec{e}_i\cdot \vec{G})^2}{|\vec{G}|^2}-\frac{1}{2}\right],
\end{equation}
 where $A_k$ is a free parameter controlling the interfacial tension, and $\vec{G}$ is the local color gradient which is defined by $\vec{G}(\vec{x},t)=\sum_i [\rho_R(\vec{x}+\vec{e}_i,t)-\rho_B(\vec{x}+\vec{e}_i,t)]\vec{e}_i$. However, Reis \& Phillips~\cite{Reis2007} and Liu et al.~\cite{Liu2012a} found that, a direct extension of the perturbation operator Eq.~(\ref{eq_d2q7}) to popular D2Q9 and D3Q19 lattices cannot recover the correct Navier-Stokes equations for two-phase flows. To obtain the correct interfacial force term for the D2Q9 lattice, Reis \& Phillips proposed an improved perturbation operator~\cite{Reis2007},
\begin{equation}\label{eq_reis}
(\Omega_i^k)^{(2)}=\frac{A_k}{2}|\vec{G}|\left[w_i\frac{(\vec{e}_i\cdot \vec{G})^2}{|\vec{G}|^2}-B_i\right],
\end{equation}
where $w_i$ is the weight factor, and $B_0=-\frac{4}{27}$,
$B_{1-4}=\frac{2}{27}$ and $B_{5-8}=\frac{5}{108}$. Using the concept of a
continuum surface force (CSF) together with the constraints of mass and
momentum conservation, a generalized perturbation operator was derived recently
by Liu et al.~\cite{Liu2012a} for the D3Q19 lattice,
\begin{equation}\label{eq_liu}
(\Omega_i^k)^{(2)}=\frac{A_k}{2}|\nabla \rho^N|\left[w_i\frac{(\vec{e}_i\cdot \nabla \rho^N)^2}{|\nabla \rho^N|^2}-B_i\right],
\end{equation}
where the phase field $\rho^N$ is defined as
\begin{equation}
\rho^N(\vec{x},t)= \frac{\rho_R(\vec{x},t) -
\rho_B(\vec{x},t)}{\rho_R(\vec{x},t) + \rho_B(\vec{x},t)},\quad
-1\leq\rho^N\leq 1,
\end{equation}
and
\begin{equation}\label{eq_Bsolution}
    B_0=-\frac{2+2\chi}{3\chi+12}c^2,\quad  B_{1-6}=\frac{\chi}{6\chi+24}c^2,\quad
    B_{7-18}=\frac{1}{6\chi+24}c^2,
\end{equation}
with $\chi$ being a free parameter. In addition, an expression for interfacial tension $\sigma$ was analytically obtained without any additional analysis and assumptions~\cite{Liu2012a},
\begin{equation}\label{eq_sigma}
\sigma=\frac{2}{9}(A_R+A_B)\tau,
\end{equation}
where $\tau$ is the relaxation time of the fluid mixture. Its validity was demonstrated by stationary bubble tests~\cite{Liu2012a}. Eq.~(\ref{eq_sigma}) suggests that the interfacial tension can be flexibly chosen by controlling $A_R$ and $A_B$.

To promote phase segregation and maintain the
interface, the recoloring operator $(\Omega_i^k)^{(3)}$ is applied, which enables the interface to be sharp, and at
the same time prevents the two fluids from mixing with each other. There are two recoloring algorithms widely used in the literature, namely the recoloring algorithm of Gunstensen et al.~\cite{Gunstensen1991} and the recoloring algorithm of Latva-Kokko and Rothman~\cite{Latva-Kokko2005}, which are hereafter referred to as A1 and A2, respectively. In A1, the distribution functions $f_i^{R \dag}(\vec{x},t)$ and $f_i^{B \dag}(\vec{x},t)$ are found by maximizing the work done by the color gradient,
\begin{equation}\label{eq_maximize}
  \sum_i [f_i^{R \dag}(\vec{x},t)-f_i^{B \dag}(\vec{x},t)]\vec{e}_i \cdot \vec{G},
\end{equation}
subject to the constraints of local conservation of the individual fluid densities of the
two components, and local conservation of the total distribution function in each
direction. This recoloring algorithm can produce a very thin interface, but generates velocity fluctuations even for a non-inclined planar
interface~\cite{Tolke2002a}. In addition, when applied to creeping flows, this recoloring algorithm can produce lattice pinning, a phenomenon where the interface can be pinned or attached to the simulation lattice rendering an effective loss of Galilean invariance~\cite{Latva-Kokko2005}. It was also identified that there is an increasing tendency for lattice pinning as both the Capillary and Reynolds numbers decrease~\cite{Halliday2006}. Therefore, this algorithm is not effective for simulating multiphase flows in porous media, especially when the capillary force is dominant. In A2, the recoloring operator is defined as~\cite{Liu2012a},
\begin{eqnarray}
&&(\Omega_i^R)^{(3)}(f_i^R)=\frac{\rho_R}{\rho}f_i^*+\beta\frac{\rho_R\rho_B}{\rho^2}\cos(\varphi_i)f_i^{eq}|_{\vec{u}=0}, \\
&&(\Omega_i^B)^{(3)}(f_i^B)=\frac{\rho_B}{\rho}f_i^*-\beta\frac{\rho_R\rho_B}{\rho^2}\cos(\varphi_i)f_i^{eq}|_{\vec{u}=0},
\end{eqnarray}
where $f_i^*$ denotes the post-perturbation, pre-segregation value
of the total distribution function along the $i$-th lattice
direction, and $f_i^{eq}=\sum_k f_i^{k,eq}$ is the total
equilibrium distribution function. $\beta$ is the segregation
parameter related to the interface thickness, and its value must
be between 0 and 1 to ensure positive particle distribution
functions. $\varphi$ is the angle between the color gradient $\vec{G}$ and the lattice vector $\vec{e}_i$, which is defined by,
\begin{equation}
\cos(\varphi_i)=\frac{\vec{e}_i \cdot
\nabla\vec{G}}{|\vec{e}_i||\nabla\vec{G}|}.
\end{equation}
Note that $\vec{G}$ should be replaced by the phase field gradient
$\nabla \rho^N$ when the perturbation operator Eq.~(\ref{eq_liu}) is applied. Leclaire et al.~\cite{Leclaire2012} conducted a numerical comparison of the recoloring operators A1 and A2 for an immiscible two-phase flow by a series of benchmark cases and concluded that the recoloring operator A2 greatly increases the rate of convergence, improves the numerical stability and accuracy of the solutions over a broad range of model parameters, and significantly reduces spurious velocities and relieves the lattice pinning problem. Several recent numerical studies~\cite{Leclaire2011,Liu2012a} indicated that, for a combination of Eq.~(\ref{eq_liu}) and the recoloring algorithm A2, the simulated density ratio and viscosity ratio can be up to $O(10^3)$ for stationary bubble/droplet tests, whereas for dynamic problems the simulated density ratio is restricted to $O(10)$ due to numerical instability.


\subsection{Inter-particle potential model}
Shan and Chen~\cite{Shan1993} developed an inter-particle potential model (also
known as Shan-Chen model) through introducing microscopic interactions among
nearest-neighboring particles. The mean field force is incorporated by using a
modified equilibrium velocity in the collision operator. This force ensures
phase separation and introduces interfacial tension. The
inter-particle potential model includes two types, namely the single-component
multiphase (SCMP) model~\cite{Shan1993,Shan1994} and the multicomponent multiphase
(MCMP) model~\cite{Shan1993,Shan1995}. In this section, we only introduce the
MCMP inter-particle potential model in the model description for the sake of
conciseness, while the capability of SCMP model and several relevant studies are
still reviewed.

The LBE for the $k$th fluid is given by,
\begin{equation}\label{eq_sc}
f_i^k(\vec{x}+\vec{e}_i\delta_t,t+\delta_t)=f_i^k(\vec{x},t)-\frac{1}{\tau_k}\left[f_i^k(\vec{x},t)-f_i^{k,eq}(\vec{x},t)\right],
\end{equation}
where the equilibrium distribution function $f_i^{k,eq}$ is written as,
\begin{equation}\label{eq_feq}
  f_i^{k,eq} = \rho_k w_i\left[1+\frac{3}{c^2}\vec{e}_i\cdot \vec{u}_k^{eq}+\frac{9}{2c^4}(\vec{e}_i\cdot
    \vec{u}_k^{eq})^2-\frac{3}{2c^2}|\vec{u}_k^{eq}|^2\right].
\end{equation}
The macroscopic density and momentum of the $k$th fluid are defined by $\rho_k = \sum_i f_i^k$ and
$\rho_k \vec{u}_k = \sum_i f_i^k\vec{e}_i$. The equilibrium velocity of the $k$th fluid is modified to carry the effect of the interactive force~\cite{Shan1995,huang:066701},
\begin{equation}\label{eq_ueq}
  \vec{u}_k^{eq} = \vec{u}'+\frac{\tau_k \vec{F}_k}{\rho_k},
\end{equation}
where $\vec{u}'$ is a common velocity, which is taken as $\vec{u}'=(\sum_k \frac{\rho_k \vec{u}_k}{\tau_k})/(\sum_k \frac{\rho_k}{\tau_k})$
to conserve the momentum in the absence of forces. $\vec{F}_k$ is the net force exerted on the $k$th fluid which includes both the fluid-fluid cohesion ($\vec{F}_k^{f-f}$) and the fluid-solid adhesion $\vec{F}_k^{f-s}$, so that $\vec{F}_k=\vec{F}_k^{f-f}+\vec{F}_k^{f-s}$.

In the inter-particle potential model, nearest neighbor interactions are used to model the fluid-fluid cohesive force~\cite{Shan1995,huang:066701},
 \begin{equation}\label{eq_ff}
   \vec{F}_k^{f-f}(\vec{x},t)=-G_c\psi_k(\vec{x},t)\sum_i w_i \psi_{\bar{k}}(\vec{x}+\vec{e}_i\delta_t,t)\vec{e}_i,
 \end{equation}
where is $G_c$ a parameter that controls the strength of the cohesive force, $k$ and $\bar{k}$
denote two different fluid components, and $\psi_k$ is the interaction
potential (or ``effective mass") which is a function of local density. Analysis
has shown that the interaction potential function has to be monotonically
increasing and bounded~\cite{Shan1993}. Several forms of the interaction
potential are commonly utilized in the literature and include, for example
$\psi_k = \rho_k$~\cite{Martys1996,huang:066701} and
$\psi_k=1-\mathrm{e}^{-\rho_k}$~\cite{Shan1993,Schmieschek2011}. The force
$\vec{F}_k^{f-f}$ allows the generation of interface between the different
fluids and the equation of state is given by $p=\frac{1}{3}c_s^2\sum_k
\rho_k+\frac{1}{6}G_c\sum_{k\bar{k}}\psi_k \psi_{\bar{k}}$~\cite{huang:066701},
where the first term corresponds to the ideal gas and the second term is the
non-ideal part.

Repulsive interactions between two components ($G_c > 0$) are utilised to model
systems of partly miscible or immiscible fluid mixtures. While the input
parameters are determined strictly phenomenologically, this approach has recently
been shown equivalent to the explicit adjustment of the free energy of
the system~\cite{PhysRevE.84.036703}.

In the context of multiphase flow in oil reservoirs, surfactants are employed
in enhanced oil recovery processes to alter the relative wettability of oil and
water. Amphiphiles (i.e. surfactants) are comprised of a hydrophilic head group
and a hydrophobic tail. Amphiphilic behaviour is modeled by a dipolar moment
$\mathbf{d}$ with orientation $\theta$ defined for each lattice site.
The relaxation is a BGK-like process, where the equilibrium moment is dependent on
the surrounding fluid densities~\cite{bib:nekovee-coveney-chen-boghosian}.  The
introduction of the dipole vector accounts for three additional Shan-Chen type
interactions, namely an additional force term,
\begin{equation}\label{eq:sccompsurfforce}
\mathbf{F}^{k, s}=-2\psi^{k} (\mathbf{x},t) G_{k,s} \sum_{i\neq 0} \mathbf{\tilde{d}}(\mathbf{x}+\mathbf{e}_i\delta_t,t)\cdot \mathbf{\Theta}_{i} \psi^{s}(\mathbf{x} +\mathbf{e}_{i}\delta_t,t),
\end{equation}
for the regular fluid components $k$ imposed by the surfactant species \emph{s}. Therein, the tilde denotes post-collision values and the second rank tensor
$\mathbf{\Theta}_{i} \equiv \mathbb{I} - 3\frac{\mathbf{e}_{i}\mathbf{e}_{i}}{\delta_x^{2}}$, with the identity operator $\mathbb{I}$ weights the dipole force contribution according to the orientation relative to the density gradient. The surfactant species is subject to forcing as well, where the contribution of the regular components \emph{$k$} is given by,

\begin{equation}\label{eq:scsurfcompforce}
\mathbf{F}^{k, s}=2\psi^{s} (\mathbf{x},t) \mathbf{\tilde{d}}(\mathbf{x},t) \cdot \sum_{k} G_{k, s} \sum_{i\neq 0} \mathbf{\Theta}_{i} \psi^{k}(\mathbf{x} +\mathbf{e}_{i}\delta_t,t),
\end{equation}
and,
\begin{eqnarray}\label{eq:scsurfsurfforce}
\mathbf{F}^{ss} = & -\frac{12}{\parallel \mathbf{c}_s \parallel^{2}} \psi^{s}(\mathbf{x},t) G_{s,s} \cdot \sum_{i\neq 0} \psi^{s}(\mathbf{x}+\mathbf{e}_{i}\delta_t,t) \cdot \bigg( \mathbf{\tilde{d}}(\mathbf{x}+\mathbf{e}_{i}\delta_t,t)\cdot \mathbf{\Theta}_{i} \\
&\cdot \mathbf{\tilde{d}}(\mathbf{x},t) \mathbf{e}_{i} 
+ \left[ \mathbf{\tilde{d}}(\mathbf{x}+\mathbf{e}_{i}\delta_t)\mathbf{\tilde{d}}(\mathbf{x},t) + \mathbf{\tilde{d}}(\mathbf{x},t)\mathbf{\tilde{d}}(\mathbf{x}+\mathbf{e}_{i}\delta_t,t) \right] \cdot \mathbf{e}_{i} \bigg),\nonumber
\end{eqnarray}
is the force due to self-interaction of the amphiphilic
species~\cite{bib:nekovee-coveney-chen-boghosian}. The amphiphilic lattice Boltzmann model has been used successfully to describe domain growth in mixtures of simple
liquids and
surfactants~\cite{bib:gonzalez-nekovee-coveney,bib:emerton-coveney-boghosian,bib:gonzalez-coveney-2},
the formation of mesophases such as the so-called primitive, diamond, and
gyroid
phases~\cite{bib:jens-gonzalez-giupponi-coveney:2005,bib:jens-giupponi-coveney:2006,bib:jens-harvey-chin-venturoli-coveney:2005,bib:saksena-coveney-2009},
and to investigate the behaviour of amphiphilic mixtures in complex geometries
such as microchannels and porous
media~\cite{bib:jens-kunert:2008b,bib:jens-kunert-jari:2010,SNH12b}.


Furthermore, a force exerted by a surface interaction can be introduced as~\cite{bib:jens-kunert-herrmann:2006,Martys1996,huang:066701},
 \begin{equation}\label{eq_fs}
   \vec{F}_k^{f-s}(\vec{x},t)=-G_{ads,k}\psi_k(\vec{x},t)\sum_i w_i s(\vec{x}+\vec{e}_i\delta_t)\vec{e}_i,
 \end{equation}
where $G_{ads,k}$ represents the strength of interaction between the fluid $k$ and the solid, and $s(\vec{x}+\vec{e}_i\delta_t)$ is an indicator function which is equal to 1 for a solid node or 0 for a fluid node, respectively. When $\psi_k$ is chosen as $\rho_k$, Huang et al.~\cite{huang:066701} proposed the following estimate for the contact angle $\theta$ (which is measured in fluid 1),
\begin{equation}\label{eq_theta_sc}
  \cos(\theta)=\frac{G_{ads,2}-G_{ads,1}}{G_c\frac{\rho_1-\rho_2}{2}},
\end{equation}
which suggests that different contact angles can be achieved by adjusting the parameters $G_{ads,k}$.

Recently, several methods have been developed to alleviate the limitations of
the original inter-particle potential model and improve its performance. These
techniques include incorporating a realistic equation of state into the
model~\cite{Yuan2006,Zhang2008}, increasing the isotropy order of the
interactive force~\cite{Shan2006,Sbragaglia2007}, improving the force scheme~\cite{Kupershtokh2009,Huang2011,Li2012a,Sun2012}, and using the
Multi-Relaxation Time (MRT) scheme~\cite{Yu2010,SNH12b} instead of the BGK approximation. These techniques have been demonstrated to be effective
in reducing the magnitude of spurious velocities, eliminating the unphysical
dependence of equilibrium density and interfacial tension on viscosity
(relaxation time), and increasing the viscosity and denstiy ratios in simple
systems~\cite{Porter2012,Bao2013,Yu2009,Kamali2012,Kamali2013}. As shown by
Porter et al.~\cite{Porter2012}, the 4th-order isotropy in the interactive force
results in stable bubble simulations for a viscosity ratio of up to 300, whereas the 10th-order isotropy resultis in stable bubble simulations for a viscosity
ratio of up to 1050. However, the effectiveness of these improved models in
dealing with multiphase flow in complex porous media has not been fully
investigated and is an active research topic. In a recent study it was found
that  the interfacial width associated with the interparticle potential model
is significantly larger than for the colour gradient model or the free-energy
model introduced below~\cite{Yang:2013}. However, this finding could not be
confirmed by the authors of the current paper. We find an interfacial width
which is comparable to the free-energy model (about five lattice units).

\subsection{Free-energy model}
The free-energy model proposed by Swift et al.~\cite{Swift1995,Swift1996} is
built upon the phase-field theory, in which a free-energy functional is used to
account for the interfacial tension effects and describe the evolution of
interface dynamics in a thermodynamically consistent manner. Similar to the
inter-particle potential model~\cite{Shan1993}, the free-energy model also
includes both SCMP model and MCMP models~\cite{Swift1996}. The SCMP
free-energy model can satisfy the local conservation of mass and momentum, but
it suffers from a lack of Galilean invariance since density (pressure)
gradients are of order $O(1)$ at liquid-gas interfaces. However, errors due to
violation of Galilean invariance are insignificant for the MCMP free-energy
model, which uses  binary fluids with similar density so that the pressure
gradients in the interfacial regions are much smaller~\cite{Sman2008}.
Therefore, the MCMP free-energy model has been applied to understand multiphase
flows in porous media especially in the situation where inertial effects can be
neglected~\cite{Hao2010,Niu2007}.

In the MCMP free-energy model for two-phase system such as fluids `1' and `2' which have density of $\rho_1$ and $\rho_2$, respectively, two distribution functions $f_i(\vec{x}, t)$ and $g_i(\vec{x}, t)$ are used to model density $\rho=\rho_1+\rho_2$, velocity $\vec{u}$ and the order parameter $\phi$ which represents the different phases, respectively. The time evolution equations for the distribution functions, using the standard BGK approximation, can be written as,
\begin{eqnarray}
\label{eq_fi}
  f_i(\vec{x}+\vec{e}_i\delta_t,t+\delta_t)-f_i(\vec{x},t) &=& \frac{1}{\tau_f}[f_i^{eq}(\vec{x},t)-f_i(\vec{x},t)], \\
  g_i(\vec{x}+\vec{e}_i\delta_t,t+\delta_t)-g_i(\vec{x},t) &=& \frac{1}{\tau_g}[g_i^{eq}(\vec{x},t)-g_i(\vec{x},t)],
  \label{eq_gi}
\end{eqnarray}
where $\tau_f$ and $\tau_g$ are two independent relaxation parameters; $f_i^{eq}$ and $g_i^{eq}$ are the equilibrium distributions of $f_i$ and $g_i$.

The underlying physical properties of lattice Boltzmann schemes are determined via the hydrodynamic moments of the
equilibrium distribution functions. The moments of the distribution functions should satisfy~\cite{Swift1996}
\begin{eqnarray}
\label{eq_fi_m1a}
  && \sum_i f_i = \sum_i f_i^{eq}=\rho; \quad \sum_i g_i = \sum_i g_i^{eq}=\phi,   \\
  && \sum_i f_i\vec{e}_{i} = \sum_i f_i^{eq} \vec{e}_i=\rho\vec{u}; \quad \sum_i g_i^{eq} \vec{e}_i=\phi\vec{u},  \\
  && \sum_i f_i\vec{e}_{i}\vec{e}_{i}^T = \vec{P}+\rho \vec{u}\vec{u}^T; \quad \sum_i g_i\vec{e}_{i}\vec{e}_{i}^T = \Gamma \mu\vec{I}+\phi \vec{u}\vec{u}^T,
\label{eq_fi_m2}
\end{eqnarray}
where $\vec{P}$ is the pressure tensor, and $\Gamma$ is a coefficient which controls the phase interface diffusion and is related to the mobility $M$ of the fluid as follows~\cite{Swift1996,Liu2009a},
\begin{equation}\label{eq_mobility}
  M=\Gamma\left(\tau_g -\frac{1}{2}\right)\delta_t.
\end{equation}

Following the constraints of Eqs.~(\ref{eq_fi_m1a})-(\ref{eq_fi_m2}), the equilibrium distributions $f_i^{eq}$ and $g_i^{eq}$, which are assumed to be a power series in terms of the local velocity, can be written as~\cite{Liu2010},
\begin{eqnarray}
  &&f_i^{eq} = F_i+\rho w_i\left(\frac{3}{c^2}\vec{e}_i\cdot \vec{u}+\frac{9}{2c^4}(\vec{e}_i\cdot
    \vec{u})^2-\frac{3}{2c^2}|\vec{u}|^2\right),\\
  && g_i^{eq} = w_i\left[\frac{\Gamma \mu}{c_s^2}+\phi\left(\frac{3}{c^2}\vec{e}_i\cdot \vec{u}+\frac{9}{2c^4}(\vec{e}_i\cdot
    \vec{u})^2-\frac{3}{2c^2}|\vec{u}|^2\right)\right],
\end{eqnarray}
for a D2Q9 lattice with $i=1,...,8$, where the coefficient $F_i$ is given by,
\begin{equation}\label{eq_gauss}
  F_i = \left\{ \begin{array}{ll}
         \vec{e}_i^T\vec{P}\,\vec{e}_i/2c^4-(P_{xx}+P_{yy})/12c^2  & \mbox{$i=1-4$},\\
         \vec{e}_i^T\vec{P}\,\vec{e}_i/8c^4-(P_{xx}+P_{yy})/6c^2 & \mbox{$i=5-8$}.\end{array} \right.
\end{equation}
In addition, the equilibrium distributions for the rest particles are chosen to ensure mass conservation, $f_0^{eq}=\rho-\sum_{i>0}f_i^{eq}$ and $g_0^{eq}=\phi-\sum_{i>0}g_i^{eq}$.

The pressure tensor $\vec{P}$ and the interfacial tension in a two-phase system, as well as the wetting boundary condition at solid walls can be derived from the free-energy functional of the system, which is defined as a function of the order parameter $\phi$ as follows~\cite{Briant2004a},
\begin{equation}\label{eq_free_energy}
  \mathscr{F}(\phi)=\int_{V}\left(\Psi(\phi)+\frac{\kappa}{2}|\nabla\phi|^2+\rho c_s^2 \ln\rho\right)dV+\int_{S}f_w(\phi_S)dS,
\end{equation}
where $\Psi(\phi)$ is the bulk free energy density and takes a double-well form, $ \Psi(\phi)=\frac{A}{4}(\phi^2-1)^2$, with $A$
being a positive constant controlling the interaction energy between two fluids. The term $\frac{\kappa}{2}|\nabla\phi|^2$
accounts for the excess free energy in the interfacial region. The surface energy density is $f_w(\phi_S)=-\omega \phi_S$, with $\phi_S$
being the order parameter on the solid surface and $\omega$ being a constant depending on the contact angle, as will be discussed later. The fluid volume and fluid wall interface are denoted as $V$ and $S$, respectively. Note that the
final term in the first integral does not affect the phase behavior, and
is introduced to enforce incompressibility in the LBM.

The chemical potential $\mu$ is defined as the variational derivative of the free energy functional with respect to the order parameter,
\begin{equation}\label{eq_mu}
  \mu=\frac{\delta \mathscr{F}}{\delta \phi}=\Psi'(\phi)-\kappa\nabla^2\phi=A\phi(\phi^2-1)-\kappa\nabla^2\phi.
\end{equation}
The pressure tensor is responsible for generation of interfacial tension, and should follow the Gibbs-Duhem relation~\cite{Liu2011a},
\begin{equation}\label{eq_Gibs}
  \nabla\cdot\vec{P}=\nabla \rho c_s^2 + \phi\nabla \mu.
\end{equation}
A suitable
choice of pressure tensor, which fulfils Eq.~(\ref{eq_Gibs}) and reduces to the usual bulk
pressure if no gradients of the order parameter are present, is~\cite{Liu2011a},
\begin{equation}\label{eq_ptensor}
  \vec{P}=[p_b-\frac{\kappa}{2}(\nabla \phi)^2-\kappa\phi\nabla^2\phi]\vec{I}+\kappa (\nabla\phi)(\nabla\phi)^T,
\end{equation}
where $p_b$ is the bulk pressure and given by $p_b=\rho c_s^2+A(-\frac{1}{2}\phi^2+\frac{3}{4}\phi^4)$.

For a flat interface with $x$ being its normal direction, the order parameter profile across the interface can be given by
$\phi=\tanh(x/\xi)$,  where $\xi$ is a measure
of the interface thickness, which is given by $\xi=\sqrt{2\kappa/A}$. The interfacial tension is evaluated according to thermodynamic theory as
$\sigma=\int_{-\infty}^{+\infty}\kappa(\frac{d\phi}{dx})^2dx=\frac{4\kappa}{3\xi}$.

Using the Chapman-Enskog multiscale analysis~\cite{Swift1996}, the evolution functions Eqs.~(\ref{eq_fi}) and (\ref{eq_gi}) can lead to the Navier-Stokes equations for a two-phase system and the Cahn-Hilliard equation for interface evolution under the low Mach number assumption,
\begin{eqnarray}
  && \frac{\partial \rho}{\partial t}+\nabla\cdot (\rho\vec{u})=0, \\
  && \frac{\partial (\rho\vec{u})}{\partial t}+\nabla\cdot (\rho\vec{u}\vec{u})=-\nabla\cdot\vec{P}+\nabla\cdot(\eta\nabla\vec{u}),\\
  && \frac{\partial \phi}{\partial t}+\nabla\cdot (\phi\vec{u})=M\nabla^2\mu,
\end{eqnarray}
where the dynamic viscosity $\eta$ is related to the relaxation time $\tau_f$ in Eq.~(\ref{eq_fi}) by $\eta=\rho(\tau_f-0.5)c_s^2\delta_t$. To account for unequal viscosities of the two fluids,  the viscosity at the phase interface can be evaluated by~\cite{Liu2011a,Graaf2006},
\begin{equation}\label{eq_vis}
  \eta(\phi)=\frac{1+\phi}{2}\eta_1+\frac{1-\phi}{2}\eta_2 \quad \mathrm{or} \quad \frac{1}{\eta(\phi)}=\frac{1+\phi}{2\eta_1}+\frac{1-\phi}{2\eta_2},
\end{equation}
where $\eta_1$ and $\eta_2$ denote the viscosities of fluid 1 and 2 with the equilibrium order parameter of 1 and $-1$, respectively.

Minimizing the free-energy functional $\mathscr{F}$ at equilibrium condition results in the following natural boundary condition at the wall~\cite{Briant2004a},
\begin{equation}\label{eq_wallfree}
  \kappa\frac{\partial \phi}{\partial n} =-\omega,
\end{equation}
where $\vec{n}$ is the local normal direction of the wall pointing into
the fluid. The static contact angle $\theta$ (measured in the fluid 1) can be shown to satisfy the following equation,
\begin{equation}\label{eq_costheta}
 \cos(\theta)=\frac{(1+\Theta)^{3/2}-(1-\Theta)^{3/2}}{2},
\end{equation}
where the wetting potential $\Theta$ is given by,
\begin{equation}\label{eq_Theta_dimen}
  \Theta=\omega/\sqrt{\kappa A/2}.
\end{equation}
From Eq.~(\ref{eq_costheta}), the wetting potential can be obtained explicitly as,
\begin{equation}\label{eq_formu}
  \Theta=2\mathrm{sign}\left(\frac{\pi}{2}-\theta\right)\left[\cos{\frac{\beta}{3}}\left(1-\cos{\frac{\beta}{3}}\right)\right]^{1/2},
\end{equation}
where $\beta=\arccos(\sin^2\theta)$ and sign(.) is the sign function.

The wetting boundary condition at the solid wall can be implemented following the method proposed by Niu et al.~\cite{Niu2007}. In this method, the order parameter derivative in Eq.~(\ref{eq_wallfree}) is evaluated by the first-order finite difference as $\partial \phi/\partial n = (\phi_f-\phi_S)/\delta_x$ with $\phi_S$ being the order parameter of the solid and $\phi_f$ the order parameter of the fluid lattices adjacent to the solid wall. By substituting the finite differences into Eq.~(\ref{eq_wallfree}) and averaging them over all fluid nodes adjacent to the solid wall, the order parameter $\phi_S$ can be approximated by,
\begin{equation}\label{eq_phis}
  \phi_S=\frac{\sum_N (\phi_f-\frac{\omega}{\kappa}\delta_x)}{N}.
\end{equation}
Here $N$ is the total number of the fluid sites which are nearest to the solid
walls. Note that Eq.~(\ref{eq_phis}) can be easily applied to complex solid
boundaries as in porous media.

In the MCMP free-energy model developed by Swift et al.~\cite{Swift1996}, which introduces the interfacial tension
force by imposing additional constraints on the equilibrium distribution function, the unphysical spurious velocities, caused by a slight imbalance between the stresses in the interfacial region, are pronounced near the interfaces
and solid surfaces. Pooley et al.~\cite{Pooley2008} identified that the strong spurious velocities in
the steady state lead to an incorrect equilibrium contact angle for  binary fluids
with different viscosities. The key to reducing spurious velocities lies in the formulation of treating the interfacial tension force~\cite{Lee2006}. Jacqmin~\cite{Jacqmin1999} suggested  the chemical potential form of the interfacial tension
force, guaranteed to generate motionless equilibrium states
without spurious velocities. Jamet et al.~\cite{Jamet2002} later showed that the chemical
potential form can ensure the correct energy transfer between the kinetic energy
and the interfacial tension energy. In the free-energy model of potential form, Eq.~(\ref{eq_Gibs}) is often rewritten as~\cite{Huang2009,Liu2011a},
\begin{equation}\label{eq_Gibs2}
  \nabla\cdot\vec{P}=\nabla \tilde{p} - \mu\nabla\phi,
\end{equation}
where $\tilde{p}=\rho c_s^2 + \phi\mu$ is the modified pressure. Once the pressure tensor is expressed as
Eq.~(\ref{eq_Gibs2}) in the Navier-Stokes equations, $\tilde{p}$ can be simply incorporated in the modified
equilibrium distribution function and the interfacial force term, $\vec{F}_S=-\mu\nabla\phi$, can be treated as an
external force in the lattice Boltzmann implementation. Following the work of Liu and Zhang~\cite{Liu2011a}, the time evolution equation for $f_i$ can be replaced by,
\begin{equation}\label{eq_fi2}
   f_i(\vec{x}+\vec{e}_i\delta_t,t+\delta_t)-f_i(\vec{x},t) = \frac{1}{\tau_f}[f_i^{eq}(\vec{x},t)-f_i(\vec{x},t)]+H_i,
\end{equation}
when the chemical potential form is employed. In order to recover the correct Navier-Stokes equations, the moments of $f_i^{eq}$ and $H_i$ should satisfy,
\begin{eqnarray}
\label{eq_fi_m1}
  && \sum_i f_i^{eq}=\rho; \quad \sum_i f_i^{eq} \vec{e}_i=\rho\vec{u}; \quad \sum_i f_i^{eq}\vec{e}_{i}\vec{e}_{i}^T = \tilde{p}\vec{I}+\rho \vec{u}\vec{u}^T, \\
  && \sum_i H_i=0; \quad \sum_i H_i \vec{e}_i=\delta_t(1-\frac{1}{2\tau_f})\vec{F}_S; \nonumber \\
  && \sum_i H_i\vec{e}_{i}\vec{e}_{i}^T = \delta_t(1-\frac{1}{2\tau_f})(\vec{u}\vec{F}_S^T+\vec{F}_S\vec{u}^T),
\end{eqnarray}
which leads to,
\begin{eqnarray}
  &&f_i^{eq} =  w_i\left[A_i+\rho\left(\frac{3}{c^2}\vec{e}_i\cdot \vec{u}+\frac{9}{2c^4}(\vec{e}_i\cdot
    \vec{u})^2-\frac{3}{2c^2}|\vec{u}|^2\right)\right],\\
  && H_i = \left(1-\frac{1}{2\tau_f}\right)w_i\left[\frac{\vec{e}_i-\vec{u}}{c_s^2}+\frac{\vec{e}_i\cdot\vec{u}}{c_s^4}\vec{e}_i\right]\cdot\vec{F}_S\delta_t,
\end{eqnarray}
where the coefficient $A_i$ is given by,
\begin{equation}\label{eq_Ai}
  A_i = \left\{ \begin{array}{ll}
         \tilde{p}/c_s^2  & \mbox{$i>0$},\\
         \left[\tilde{p}-(1-w_0)\tilde{p}/c_s^2\right]/w_0 & \mbox{$i=0$}.\end{array} \right.
\end{equation}



Note that the fluid velocity is re-defined as $\rho\vec{u}=\sum_i f_i\vec{e}_i+\frac{1}{2}\vec{F}_S\delta_t$ to carry some effects of the external force. Although the free-energy model proposed by Swift et
al.~\cite{Swift1996} and its potential form are completely equivalent mathematically, they produce
different discretization errors for the calculation of the interfacial tension force, leading to the difference in magnitude of spurious velocities.
It can be observed that the free-energy model of potential form is able to produce much smaller spurious velocity than the other two models due to the smaller discretization error introduced in the treatment of interfacial tension force. Since spurious velocities are effectively suppressed, the free-energy model of potential form with SRT and bounce-back boundary condition is also capable of capturing the correct equilibrium contact angle for both fluids with different viscosities.

\subsection{Mean-field theory model}
In the mean-field theory model~\cite{He1999},  interfacial dynamics, such as phase segregation
and interfacial tension, are modeled by incorporating molecular interactions.
Using the mean-field approximation for intermolecular interaction and following the treatment of the excluded volume effect by Enskog, the effective intermolecular force can be expressed as,
\begin{equation}\label{eq_intermolecule}
  \vec{F}=-\nabla\psi+\vec{F}_S =-\nabla\psi+\kappa\rho\nabla\nabla^2\rho,
\end{equation}
where $\psi(\rho)$ is a function of the density, and is related to the pressure by $\psi(\rho)=p - \rho c_s^2$. The pressure $p$ is chosen to satisfy the Carnahan-Starling equation of state,
\begin{equation}\label{eq_CS}
  p(\phi)=\phi RT\left[\frac{1+\phi+\phi^2-\phi^3}{(1-\phi)^3}\right]-a\phi^2,
\end{equation}
where $R$ is the gas constant, $T$ is the temperature, and the
parameter $a$ determines the attraction strength.

The lattice Boltzmann equations are derived from the continuous Boltzmann equation with
appropriate approximations suitable for incompressible flow. The
stability is improved by reducing the effect of numerical errors in calculation
of molecular interactions. Specifically, two distribution functions, an index distribution
function $f_i$ and a pressure distribution function $g_i$, are
employed to describe the evolution of the order parameter
and the velocity/pressure field, respectively, and the LBEs for
the two distributions are~\cite{He1999},
\begin{eqnarray}
\label{eq_he_fi}
 && f_i(\vec{x}+\vec{e}_i\delta_t,t+\delta_t)-f_i(\vec{x},t) = -\frac{f_i(\vec{x},t)-f_i^{eq}(\vec{x},t)}{\tau} \nonumber \\
 &&-\left(1-\frac{1}{2\tau}\right)\frac{(\vec{e}_i-\vec{u})\cdot\nabla\psi(\phi)}{c_s^2}\Gamma_i(\vec{u})\delta_t, \\
  && g_i(\vec{x}+\vec{e}_i\delta_t,t+\delta_t)-g_i(\vec{x},t) = -\frac{g_i(\vec{x},t)-g_i^{eq}(\vec{x},t)}{\tau} \nonumber \\
 &&-\left(1-\frac{1}{2\tau}\right)\left(\vec{e}_i-\vec{u}\right)\cdot\{\Gamma_i(\vec{u})\vec{F}_S-(\Gamma_i(\vec{u})-\Gamma_i(\vec{0}))\nabla\psi(\rho)\}\delta_t,
\label{eq_he_gi}
\end{eqnarray}
where $\tau$ is the relaxation time that is related to
the kinematic viscosity by $\nu=c_s^2(\tau-0.5)\delta_t$ and the function $\Gamma_i$ is defined by,
\begin{equation}\label{eq_gamma_i}
  \Gamma_i(\vec{u})=w_i\left(1+\frac{\vec{e}_i\cdot \vec{u}}{c_s^2}+\frac{(\vec{e}_i\cdot
    \vec{u})^2}{2c_s^4}-\frac{|\vec{u}|^2}{2c_s^2}\right).
\end{equation}
The equilibrium distribution functions $f_i^{eq}$ and $g_i^{eq}$ are taken as,
\begin{eqnarray}
  f_i^{eq} &=& w_i \phi\left[1+\frac{\vec{e}_i\cdot \vec{u}}{c_s^2}+\frac{(\vec{e}_i\cdot
    \vec{u})^2}{2c_s^4}-\frac{|\vec{u}|^2}{2c_s^2}\right], \\
  g_i^{eq} &=& w_i \left\{p+\rho c_s^2\left[\frac{\vec{e}_i\cdot \vec{u}}{c_s^2}+\frac{(\vec{e}_i\cdot
    \vec{u})^2}{2c_s^4}-\frac{|\vec{u}|^2}{2c_s^2}\right]\right\}.
\end{eqnarray}

The macroscopic variables are calculated through,
\begin{equation}\label{eq_macros_he}
  \phi = \sum_i f_i; \quad p = \sum_i g_i -\frac{1}{2}\vec{u}\nabla \psi(\rho)\delta_t; \quad \rho\vec{u}=\frac{1}{c_s^2}\sum_i g_i \vec{e}_i + \frac{1}{2}\vec{F}_S\delta_t.
\end{equation}
The density and kinematic viscosity of the fluid mixture are
calculated from the index function,
\begin{equation}
  \rho(\phi)=\rho_V+\frac{\phi-\phi_V}{\phi_L-\phi_V}(\rho_L-\rho_V); \quad  \nu(\phi)=\nu_V+\frac{\phi-\phi_V}{\phi_L-\phi_V}(\nu_L-\nu_V),
\end{equation}
where $\rho_L$ and $\rho_V$ are the densities of the liquid and vapor
phase, respectively, $\nu_L$ and $\nu_V$ are the corresponding kinematic viscosities,
and $\phi_L$ and $\phi_V$ are the minimum and maximum values of the
index function, respectively, which can be obtained through Maxwell's equal area construction. For $a = 12RT$, one can obtain $ \phi_G= 0.02283$ and $\phi_L =0.25029$. By the transformation of the particle distribution
function for mass and momentum into that for hydrodynamic pressure and momentum, the numerical stability is enhanced in Eq.~(\ref{eq_he_gi}) due to reduction of discretization error of the forcing term (i.e. the leading order of
the intermolecular forcing terms was reduced from $O(1)$ to $O(\vec{u})$~\cite{He1999}). Although this model is more robust than most of the previous LBMs, it is restricted to density ratios up to approximately 15~\cite{Fakhari2010}. The derivation and more details of the mean-field theory model can be found in~\cite{He1999}. In the mean-field theory model, the interfacial tension is controlled by the parameter $\kappa$ in $\vec{F}_S$, which plays a similar role as the interaction parameter $G_c$ in the inter-particle potential model, and therefore stationary bubble tests are required to obtain the value of interfacial tension in practical applications. In addition, in order to introduce wetting properties at the solid surface, Yiotis et al.~\cite{Yiotis2007} proposed imposing $\rho=\rho_S$ on lattice nodes within the solid phase. By choosing $\rho_S$, different contact angles can be achieved on the solid surface.

\subsection{Stabilized diffuse-interface model}
Lattice Boltzmann simulation of multiphase flows with high density ratios (HDRs) is a challenging task~\cite{Meakin2009}. There has been an ongoing effort to improve the stability of LBM for HDR multiphase
flows. To date, the most commonly used  HDR multiphase LBMs include the free-energy approach of Inamuro et al.~\cite{Inamuro2004}, the HDR model of Lee \& Lin~\cite{Lee2005}, and the stabilized diffuse-interface model~\cite{Lee2010}. However, the former two have exposed some deficiencies. In the free-energy approach of Inamuro et al.~\cite{Inamuro2004}, a projection step is applied to enforce the continuity condition after every collision-stream step, which would reduce greatly the efficiency of the method. Also, this approach needs to specify the cut-off value of the order parameter in order to avoid numerical instability, which can give rise to some non-physical disturbances even though the divergence of the velocity field is zero, and it is therefore inaccurate for many incompressible flows although the projection step is employed to secure the incompressible condition. As pointed out by Zheng et al.~\cite{Zheng2006}, the HDR model of Lee \& Lin~\cite{Lee2005} cannot lead to the correct governing equation for interface evolution (i.e. the Cahn-Hilliard equation). In addition, some additional efforts are still required for this model to account for the wetting of fluid-solid interfaces. The stabilized diffuse-interface model has great potential to simulate HDR multiphase flows at the pore scale in porous media, and can simulate a density ratio as high as 1000 with negligible spurious velocities and correctly model contact-line dynamics. Essentially, this model possesses an identical theoretical basis (i.e. Cahn-Hilliard theory) with the CFD-based phase-field (PF) method. It can be regarded as the PF method solved by the LBEs with a stable discretization technique~\cite{Liu2013}.

In the stabilized diffuse-interface model, the two-phase fluids, e.g., a gas and liquid, are assumed to be incompressible,
immiscible, and have different densities and viscosities. The order
parameter $C$ is defined as the volume fraction of one of the two phases. Thus, $C$ is assumed to be constant in the bulk phases (e.g. $C = 0$ for the gas phase while $C = 1$ for the liquid phase). Assuming that interactions between the fluids and the solid
surface are of short-range and appear in a surface integral, the
total free energy of a system is taken as the following form~\cite{Lee2010},
\begin{equation}\label{eq_15}
    \Psi_b + \Psi_s = \int_V\left(E_0(C)+\frac{\kappa}{2}|\nabla
    C|^2\right)dV  + \int_S\left(\phi_0-\phi_1C_s+\phi_2C_s^2-\phi_3C_s^3+\cdots\right)dS,
\end{equation}
where the bulk energy is taken as
$E_0=\beta C^2 (1-C)^2$ with $\beta$ being a constant, $\kappa$ is
the gradient parameter, $C_s$ is the order parameter at a solid
surface, and $\phi_i$ with $i = 0, 1, 2,\cdots$ are constant
coefficients.
The chemical potential $\mu$ is defined as the variational
derivative of the volume-integral term in Eq.~(\ref{eq_15}) with
respect to $C$,
\begin{equation}\label{eq_16}
   \mu=\frac{\partial E_0}{\partial C}-\kappa \nabla^2 \phi=2\beta
   C(C-1)(2C-1)-\kappa\nabla^2 \phi.
\end{equation}
For a planar interface at equilibrium, the interfacial profile can
be obtained through $\mu=0$,
\begin{equation}\label{eq_17}
    C(x)=\frac{1}{2}+\frac{1}{2}\tanh\left(\frac{2x}{\xi}\right),
\end{equation}
where $\xi$ is the interface thickness defined by,
\begin{equation}\label{eq_xi_hdr}
  \xi=\sqrt{8\kappa/\beta}.
\end{equation}
The interfacial tension between liquid and gas is defined as the
excess of free energy at the interface,
\begin{equation}\label{eq_18}
    \sigma=\int_0^1\sqrt{2\kappa E_0(C)}dC=\frac{\sqrt{2\kappa \beta}}{6}.
\end{equation}
Eqs.~(\ref{eq_xi_hdr}) and (\ref{eq_18}) suggest that the interfacial
tension and the interface thickness are easily controlled through the parameters
$\kappa$ and $\beta$.

In order to prevent the negative equilibrium density on a
non-wetting surface, it is necessary to retain the higher-order
terms in $\Psi_S$. By choosing $\phi_0=\phi_1=0$,
$\phi_2=\frac{1}{2}\phi_c$, and $\phi_3=\frac{1}{3}\phi_c$, a
cubic boundary condition for $\nabla^2 C$ is established~\cite{Lee2010},
\begin{equation}\label{eq_19}
    \frac{\partial C}{\partial n}|_s=\frac{\phi_c}{\kappa}\left(C_s-C_s^2\right),
\end{equation}
where $\phi_c$ is related to the equilibrium contact angle
$\theta$ via Young's law,
\begin{equation}\label{eq_20}
    \phi_c=-\sqrt{2\kappa\beta}\cos(\theta).
\end{equation}
Note that the cubic boundary condition has been widely used to simulate two-phase flows
with moving contact lines~\cite{JACQMIN2000,KHATAVKAR2007a,Wiklunda2011,liu2014}. It was demonstrated numerically
that such a boundary condition can eliminate the spurious variation of
the order parameter at solid boundaries, thereby facilitating the better capturing of the correct
physics than its lower-order counterparts (e.g. Eq.~(\ref{eq_wallfree}))~\cite{Wiklunda2011}.

Considering the second order derivative term of the chemical potential
in the Cahn-Hilliard equation, a zero-flux boundary condition should be imposed at the solid boundary to ensure no diffuse flux across the boundary,
\begin{equation}\label{eq_21}
    \frac{\partial \mu}{\partial n}|_S=0.
\end{equation}

Similar to the mean-field theory model, two particle distribution functions (PDFs) are employed in the stabilized diffuse-interface model. One is the order parameter distribution function, which is used to capture the interface between different phases, and the other is the pressure distribution function for solving the hydrodynamic pressure and fluid momentum. The evolution equations of the PDFs can be derived through the
discrete Boltzmann equation (DBE) with the trapezoidal rule applied
along characteristics over the time interval ($t,t+\delta_t$)~\cite{Lee2010}. To ensure numerical stability in
solving HDR problems, the second-order biased difference scheme is applied
to discretize the gradient operators involved in forcing terms at the time $t$ while the standard central difference scheme
is applied at the time $t+\delta_t$~\cite{Lee2005,Lee2010}. The resulting evolution equations are~\cite{Liu2013},
\begin{eqnarray}
\label{eq_gia}
  &g_{\alpha}(\vec{x}+\vec{e}_{\alpha}\delta_t,t+\delta_t)-g_{\alpha}(\vec{x},t)=\frac{1}{\tau+1/2}\left[g_{\alpha}^{eq}(\vec{x},t)-g_{\alpha}(\vec{x},t)\right] \nonumber \\
   &\left.+ \delta_t(\vec{e}_{\alpha}-\vec{u})\cdot\left[\nabla^{\mathrm{MD}}\rho
   c_s^2(\Gamma_{\alpha}-w_{\alpha})-(C\nabla^{\mathrm{MD}}\mu-\rho \vec{g})\Gamma_{\alpha}\right]\right|_{(\vec{x},t)}, \\
   &h_{\alpha}(\vec{x}+\vec{e}_{\alpha}\delta_t,t+\delta_t)=h_{\alpha}^{eq}(\vec{x},t)+\frac{\delta_t}{2}M_c\nabla^2\mu\Gamma_{\alpha}|_{(\vec{x},t)}+\frac{\delta_t}{2}M_c\nabla^2\mu\Gamma_{\alpha}|_{(\vec{x}+\vec{e}_{\alpha}\delta_t,t)} \nonumber \\
   &\left.+ \delta_t(\vec{e}_{\alpha}-\vec{u})\cdot\left[\nabla^{\mathrm{MD}}C-\frac{C}{\rho c_s^2}(\nabla^{\mathrm{MD}} p+C\nabla^{\mathrm{MD}} \mu-\rho \vec{g})\right]\Gamma_{\alpha}\right|_{(\vec{x},t)},
\label{eq_hi}
\end{eqnarray}
where $\vec{g}$ is the gravitational acceleration, $g_{\alpha}$ and $h_{\alpha}$ are the PDFs for the momentum
and the order parameter, respectively, and $g_{\alpha}^{eq}$ and
$h_{\alpha}^{eq}$ are the corresponding equilibrium PDFs. The superscript `MD' on gradient denotes the
second-order mixed difference, which is an average of the central difference (denoted by the superscript `CD') and the biased difference (denoted by the superscript `BD'). As suggested in Ref.~\cite{Lee2010}, the
directional derivatives (of a variable $\varphi$) are evaluated by,
\begin{eqnarray}
  \left. \delta_t \vec{e}_{\alpha}\cdot\nabla^{\mathrm{CD}}\varphi \right|_{(\vec{x})} &=& \frac{1}{2}\left[\varphi(\vec{x}+\vec{e}_{\alpha}\delta_t)-\varphi(\vec{x}-\vec{e}_{\alpha}\delta_t)\right], \\
  \label{eq_BD_dire}
  \left. \delta_t \vec{e}_{\alpha}\cdot\nabla^{\mathrm{BD}}\varphi \right|_{(\vec{x})} &=& \frac{1}{2}\left[4\varphi(\vec{x}+\vec{e}_{\alpha}\delta_t)-\varphi(\vec{x}+2\vec{e}_{\alpha}\delta_t)-3\varphi(\vec{x})) \right], \\
 \left. \delta_t \vec{e}_{\alpha}\cdot\nabla^{\mathrm{MD}}\varphi \right|_{(\vec{x})} &=& \frac{1}{2}\left(\left. \delta_t \vec{e}_{\alpha}\cdot\nabla^{\mathrm{CD}}\varphi \right|_{(\vec{x})}+\left. \delta_t \vec{e}_{\alpha}\cdot\nabla^{\mathrm{BD}}\varphi \right|_{(\vec{x})}\right).
\end{eqnarray}
Derivatives other than the directional derivatives can be obtained by taking moments of the directional derivatives with appropriate weights. The first- and second-order derivatives are calculated as~\cite{Lee2010},
\begin{eqnarray}
\label{eq_gradCD}
  \left.\nabla^{\mathrm{CD}}\varphi\right|_{(\vec{x})} &=& \frac{1}{2c_s^2\delta_t}\sum_{\alpha} w_{\alpha}\vec{e}_{\alpha}\left[\varphi(\vec{x}+\vec{e}_{\alpha}\delta_t)-\varphi(\vec{x}-\vec{e}_{\alpha}\delta_t)\right], \\
  \label{eq_gradBD}
  \left.\nabla^{\mathrm{BD}}\varphi\right|_{(\vec{x})} &=& \frac{1}{2c_s^2\delta_t}\sum_{\alpha} w_{\alpha}\vec{e}_{\alpha}\left[4\varphi(\vec{x}+\vec{e}_{\alpha}\delta_t)-\varphi(\vec{x}+2\vec{e}_{\alpha}\delta_t)-3\varphi(\vec{x})) \right], \\
  \left.\nabla^{\mathrm{MD}}\varphi\right|_{(\vec{x})} &=& \frac{1}{2}\left(\left.\nabla^{\mathrm{CD}}\varphi\right|_{(\vec{x})}+\left.\nabla^{\mathrm{BD}}\varphi\right|_{(\vec{x})}\right), \\
  \label{eq_lap}
  \left.\nabla^{2}\varphi\right|_{(\vec{x})} &=& \frac{1}{c_s^2\delta_t^2}\sum_{\alpha} w_{\alpha}\left[\varphi(\vec{x}+\vec{e}_{\alpha}\delta_t)-2\varphi(\vec{x})+\varphi(\vec{x}-\vec{e}_{\alpha}\delta_t)\right].
\end{eqnarray}
The equilibrium PDFs $g_{\alpha}^{eq}$ and $h_{\alpha}^{eq}$ are given by,
\begin{equation}\label{eq_geq}
g_{\alpha}^{eq}=w_{\alpha}(p-\rho c_s^2)+\rho
c_s^2\Gamma_{\alpha}-\frac{\delta_t}{2}(\vec{e}_{\alpha}-\vec{u})\cdot\left[\nabla^{\mathrm{CD}}\rho
c_s^2(\Gamma_{\alpha}-w_{\alpha})-(C\nabla^{\mathrm{CD}}\mu-\rho \vec{g})\Gamma_{\alpha}\right],
\end{equation}
\begin{equation}\label{eq_heq}
h_{\alpha}^{eq}=C\Gamma_{\alpha}-\frac{\delta_t}{2}(\vec{e}_{\alpha}-\vec{u})\cdot\left[\nabla^{\mathrm{CD}}C-\frac{C}{\rho
c_s^2}(\nabla^{\mathrm{CD}} p+C\nabla^{\mathrm{CD}} \mu-\rho \vec{g})\right]\Gamma_{\alpha}.
\end{equation}
Through the Chapman-Enskog analysis~\cite{Li2012,Liu2013a},
the following macroscopic equations can be derived from
Eqs.(\ref{eq_gia}) and (\ref{eq_hi}) in the low Mach number limit,
\begin{eqnarray}
  && \frac{1}{\rho c_s^2}\left(\frac{\partial p}{\partial t}+\vec{u}\cdot \nabla p \right)+\nabla\cdot \vec{u}=0, \\
  \label{eq_mom_hdr}
  && \frac{\partial (\rho\vec{u})}{\partial t}+\nabla\cdot (\rho\vec{u}\vec{u})=-\nabla p+\nabla\cdot\left[\eta(\nabla\vec{u}+\nabla\vec{u}^T)\right]-C\nabla \mu +\rho\vec{g},\\
  && \frac{\partial \phi}{\partial t}+\vec{u}\cdot\nabla\phi=M\nabla^2\mu,
\end{eqnarray}
where the dynamic viscosity is given by $\eta=\rho\tau c_s^2\delta_t$. For incompressible flows, $\partial_t p$ is negligibly small and $\vec{u}\cdot\nabla p$ is of the order of $O(Ma^3)$. Thus,
the divergence-free condition can be approximately satisfied.
However, Eq.~(\ref{eq_mom_hdr}) is inconsistent with the target momentum equation
in the phase-field model due to the error term $\vec{u}(\partial_t \rho +\vec{u}\cdot\nabla \rho)\neq \vec{0}$. To eliminate the error term and recover the correct momentum equation, Li et al.~\cite{Li2012} and Liu et al.~\cite{Liu2013a} proposed to introduce an additional force term, $\frac{d\rho}{d C}M\nabla^2 \mu\vec{u}$ to the LBEs. Considering that the Reynolds number is typically very small, the additional force term is believed to have only a slight effect on  multiphase flows in porous media~\cite{Li2012}. Hence, the additional force term is not involved in the above evolution equations of PDFs for the sake of simplicity.

Finally, the order parameter, the hydrodynamic pressure and the fluid velocity are calculated by taking the zeroth- and
the first-order moments of the PDFs,
\begin{equation}\label{eq_macros}
   C=\sum_{\alpha}h_{\alpha}, ~
   \rho\vec{u}=\frac{1}{c_s^2}\sum_{\alpha}g_{\alpha}\vec{e}_{\alpha}-\frac{\delta_t}{2}(C\nabla^{\mathrm{CD}}\mu-\rho \vec{g}),
   ~
   p=\sum_{\alpha}g_{\alpha}+\frac{\delta_t}{2}\vec{u}\cdot\nabla^{\mathrm{CD}}\rho
   c_s^2.
\end{equation}
and the density and the relaxation time of the fluid mixture are calculated by~\cite{Lee2010},
\begin{eqnarray}
  \rho & =& \rho_L C+ \rho_G(1-C),\\
  \frac{1}{\tau}&=&\frac{C}{\tau_L} +\frac{1-C}{\tau_G},
  \label{eq_tau_hdr}
\end{eqnarray}
where $\tau_L$ ($\tau_G$) is the relaxation time of liquid (gas) phase. It was shown~\cite{Lee2010} that Eq.~(\ref{eq_tau_hdr}) can produce 
monotonically  varying dynamic viscosity, whereas a popular choice with $\tau$ calculated by $\tau=\tau_L C+ \tau_G(1-C)$  shows a peak of dynamic viscosity in the interface region with a magnitude several times larger than the bulk viscosities.

This model's capability for HDR multiphase flows has been validated by several benchmark cases including the test of Laplace's law, simulation of static contact angles, as well as droplet deformation and breakup in a simple shear flow~\cite{Lee2010,Zheng2013}. It was found that this model can simulate two-phase flows with a liquid to gas density ratio approaching 1000. An addition, spurious velocities produced by the model are small, which is attributed to the interfacial force of potential form and the stable numerical discretization for estimating various derivatives. However, compared with other multiphase LBMs, the stabilized diffuse-interface model is quite complex and the computational efficiency is very low since the
numerical implementation involves the discretization of many directional derivatives which need to be evaluated in every
lattice direction. Liu et al.~\cite{Liu2012a} recently presented a quantitative comparison of the required
computational time between the color gradient model and the stabilized diffuse-interface
model. Both models were  used to simulate the stationary bubble case with a
density ratio of 100. The required CPU time per timestep is roughly twice as long for the stabilized
diffuse-interface model as compared to the color gradient model.
In addition, the stabilized
diffuse-interface model needs 23 times more timesteps to achieve the same
stopping criterion. Similar to the free-energy model, this model is also built
upon the phase-field theory, so that small droplets/bubbles also tend to
dissolve as the system evolves towards an equilibrium state. Previous numerical
experiments have demonstrated~\cite{Zhang2010,Liu2013a} that a feasible
approach for reducing the droplet dissolution is to replace the constant
mobility with a variable one, which depends on the order parameter through, for
example, $M=M_c\sqrt{C^2(1-C)^2}$ with $M_c$ being a constant.

\subsection{Particle suspensions}
\label{Sec:particles}
The terms ``multiphase'' or ``multicomponent'' flow might not only describe mixtures of different fluids or fluid phases, but are also adequate to classify
fluid flows with floating objects such as suspensions or polymer solutions. In porous media applications suspension flows are relevant in the context of, for example, underground transport of liberated sand, clay or contaminants, filter
applications, or the development of highly efficient catalysts.  The
individual particles are usually treated by a particle-based method, such as the discrete element method (DEM) or molecular dynamics (MD), and momentum is transferred between them and the fluid after a sufficiently small number of timesteps.

The available coupling algorithms can be distinguished in
two classes. If the particles are smaller than the lattice Boltzmann
grid spacing, they can be treated as point particles exchanging a Stokes drag force and eventually friction forces with the fluid. This so-called friction coupling was first introduced by Ahlrichs and
D\"unweg and became particularly popular for the simulation
of polymers made of bead-spring chains or compound
particles~\cite{Ahlrichs99,Ahlrichs01,Lobaskin04b}.

If the hydrodynamic flow around the individual particles becomes important, particles are generally discretized on the LBM lattice and at every discretization point, the local momentum exchange between particle and
fluid is computed at every timestep. This method was pioneered by Ladd and colleagues and is mostly used for suspension flows~\cite{Ladd94a,Ladd94b,Ladd01,bib:nguyen-ladd-02}.
The method has been applied to suspensions of spherical and non-spherical particles by various
authors~\cite{bib:jfm.CAiYLuEDi.1998,bib:jens-komnik-herrmann:2004,bib:jens-janoschek-toschi-2010b}.
Recently, it has been extended to particle suspensions involving multiple fluid
components~\cite{bib:jens-jansen-2011,bib:jens-frijters-2012a,bib:stratford-adhikari-pagonabarraga-desplat-cates-2005,bib:joshi-sun}.

The coupling of particles to the LBM can also be achieved through an immersed moving boundary (IMB) scheme~\cite{Noble1998,Cook2004,Owen2011}. This sub-grid-scale condition maintains the locality of LBM computations, addresses the momentum discontinuity of binary bounce back schemes and provides reasonable accuracy for obstacles mapped at low resolution.

To simulate the hydrodynamic interactions between solid particles in
suspensions, the lattice Boltzmann model has to be modified to incorporate the boundary conditions imposed on the fluid by the solid particles. Stationary solid objects are introduced into the model by replacing the usual collision rules at a specified
set of boundary nodes by the ``link-bounce-back'' collision rule~\cite{bib:nguyen-ladd-02}.  When placed on the lattice, the boundary surface cuts some of the links between lattice nodes. The fluid particles moving along these links interact with the solid surface at boundary nodes placed halfway along the links. Thus, a discrete representation of the surface is obtained, which becomes more and more precise as the surface curvature gets smaller and which is exact for surfaces parallel to lattice planes. Since the velocities in the LBM are discrete, boundary conditions for moving suspended particles cannot be implemented directly. Instead, one modifies the density of returning particles in a way that the momentum transferred to the solid is the same as in the continuous velocity case. This is implemented by introducing an additional term,
\begin{equation}\label{eq:moving-collision-rule}
\Delta_{b,i}=\frac{2\omega^{c_i}\rho\vec{u}_b\cdot\vec{c}_i}{c_s^2}
\end{equation}
in the discrete Boltzmann equation~\cite{Ladd94a},
with $\vec{u}_b$ being the velocity of the boundary.
To avoid redistributing fluid mass from lattice nodes being covered or
uncovered by solids, one can allow interior fluid within closed surfaces. Its movement relaxes to the movement of the solid body on much shorter time scales than the characteristic hydrodynamic interaction~\cite{Ladd94a,bib:nguyen-ladd-02,bib:heemels-hagen-lowe}.



\section{Implementation Strategies}
\label{sec:imp}
A number of features of the LBM facilitate straightforward distribution on massively parallel systems \cite{Satofuka1999}. In particular, the method is typically implemented on a regular, orthogonal grid, and the collision operator and many boundary implementations are local processes meaning that each lattice node only requires information from its own location to be relaxed. However, it should be noted that some extensions of the method require the calculation of velocity and strain gradients from non-local information, and this complicates parallelization somewhat.

Given the current state of computational hardware, in particular the relative speed and capacity of processors and memory, the LBM is a memory-bound numerical method. This means that the time required to read and write data from and to memory, not floating point operations, is the critical bottleneck to performance. This has a number of implications for the implementation of the method, be it on shared-memory multicore nodes, distributed memory clusters, or graphical processing units (GPUs). Each of these parallelization strategies is discussed as follows.

\subsection{Pore-list versus pore-matrix implementations}
In typical lattice Boltzmann codes used for the simulation of flow in porous
media, the pore space and the solid nodes are represented by an array including
the distribution functions $f_i$ and a Boolean variable to distinguish between
a pore and a matrix node (``pore-matrix'' or ``direct addressing''
implementation). At every timestep the loop covering the domain includes the
fluid and the solid nodes and if-statements are used to distinguish whether the
collision and streaming steps or bounary conditions need to be applied. The
advantage of this data structure is its straightforward implementation.
However, for the simulation of fluid flow in porous media with low porosity the
drawbacks are high memory demands and inefficient loops through the whole
simulation domain~\cite{bib:jens-narvaez:2010}.

Alternatively, a data structure known as ``pore-list'' or ``indirect addressing'' can be used~\cite{2005ASIM:DXHHW}. Here, the array comprising the lattice structure contains the position (pore-position-list) and connectivity (pore-neighbor-list) of the fluid nodes only. It can be generated from the original lattice before the first timestep of the simulation. Then, only loops through the list of pore nodes not comprising any if-statements for the lattice Boltzmann algorithm are required. The CPU time needed to generate and save the
pore-list data is comparable to the computational time required for a single timestep of the usual lattice Boltzmann algorithm based on the pore-matrix data structure. This alternative approach is slightly more complicated to implement, but allows highly efficient simulations of flows in geometries with a low porosity. If the porosity becomes too large, however, the additional overhead due to the connection matrix reduces the benefits and at some point renders the method less efficient than a standard implementation. For representative 3D simulation codes it was found that the transition porosity where one of the two implementations becomes more efficient is around
40\%~\cite{bib:jens-narvaez:2010}. In addition, if the microstructure of a porous medium is not static,
but evolving due to processes like dissolution/precipitation~\cite{Chen2013}, the operation to generate and save the
pore-list data needs to be included in the time loop. In this case, the ``pore-matrix'' or ``direct addressing''
implementation will be preferred.

Ma et al.~\cite{Ma:2010} have proposed the SHIFT algorithm where the
distribution functions and the geometry of the porous medium are stored in a
single array following the ``pore-list'' idea. Smart arrangement of the data in
one-dimensional arrays allows to implement highly optimized and efficient
codes making use of the vectorization capabilities of modern CPUs.   

\subsection{Asynchronous parallelization on shared-memory, multicore nodes}
\label{sec:multicore}
Historically, the parallel processing of numerical methods utilized a distributed memory cluster as the underlying hardware. In this approach the computational domain is decomposed into the same number of sub-domains as there are nodes available in the cluster. A single sub-domain is processed on each cluster node at each time step and when all sub-domains have been processed, global solution data is synchronized.

The synchronization of solution data requires the creation of, and communication between, domain \textit{ghost} regions. These regions correspond to neighboring sections of the problem domain which are stored in memory on other cluster nodes but are required on a cluster node for the processing of its own sub-domain. In the LBM this is typically a 'layer' of grid points that encapsulates the local sub-domain. As a consequence of Amdahl’s Law, this can significantly degrade the scalability of the implementation.

Another challenge with distributed memory parallelism can be sub-optimal load balancing, which also degrades parallel efficiency, however some strategies to address this problem are discussed in Sec.~\ref{sec:clusters}.

The issues of data communication and load balancing in parallel LBM implementations can be addressed by employing shared-memory, multicore hardware, fine-grained domain decomposition, and asynchronous task distribution. Access to a single memory address space removes the need for ghost regions and the subsequent transfer of data over comparatively slow node connections. Instead, all data is accessible from either local caches or global memory. Access times for these data stores are many orders of magnitude shorter than cross-machine communication \cite{Pohl2003} and when used with an optimum cache-blocking strategy can significantly reduce the latency associated with data reads and writes.

Cache-blocking in this LBM implementation is optimized by utilizing fine-grained domain decomposition. Instead of partitioning the domain into one sub-domain per core, a collection of significantly smaller sub-domains is created. These sub-domains, or computational tasks, are sized to fit in the low-level cache of a processing core, which minimizes the time spent reading and writing data as a task is processed. In the LBM, cubic nodal bundles are used to realise fine-grained domain decomposition and on a multicore server with a core count on the order of $10^1$ the number of tasks could be in the order of $10^4$.

Parallel distribution of computational tasks requires the use of a coordination tool to manage them onto processing cores in a load balanced way. Instead of using a traditional scatter-gather approach, here the H-Dispatch distribution model \cite{Holmes2010} is used because of the demonstrated advantages for performance and memory efficiency. Rather than scatter or push tasks from the domain data structure to threads, here threads request tasks when free. H-Dispatch manages these asynchronous requests using event handlers and distributes tasks to the requesting threads accordingly. When all tasks in the problem space have been dispatched and processed, H-Dispatch identifies step completion and the process can begin again. By using many more tasks than cores, and events-based distribution of these tasks, the computational workload of the numerical method is naturally balanced.

The shared-memory aspect of this implementation requires the consideration of race conditions. Conveniently, this can be addressed in the LBM by storing two copies of the particle distribution functions at each node (which is often done anyway) and using a \textit{pull} rather than \textit{push} streaming operation. In the pull-collide sequence, incoming functions are read from neighbor nodes (non-local read) and collided, and then written to the future set of functions for the current node (local write). On cache-sensitive multicore hardware, this sequence of operations outperforms collide-push, which requires local reads and non-local writes \cite{Pohl2003}.

The benefit of optimized cache blocking is found by varying the bundle size and
measuring the speed-up of the implementation. For example in
Ref.~\cite{Leonardi2011} and for a simple $200^3$ problem, the optimal
performance point (92\% speed-up efficiency) was found at a side length of
20~\cite{Leonardi2011}. Additionally, it was found that this optimal side
length could be applied to larger domains and still yield maximum speed-up
efficiency. This suggests that the optimum bundle size for the LBM can be
determined in an \textit{a priori} fashion for specific hardware.

\subsection{Synchronous parallelization on distributed memory clusters}
\label{sec:clusters}
A number of well established and highly scalable multiphase lattice Boltzmann implementations exist. A very limited list of examples highlighting possible implementation differences includes {\it Ludwig}~\cite{bib:ludwig}, {\it LB3D}~\cite{bib:jens-harvey-chin-venturoli-coveney:2005},
{\it walBerla}~\cite{bib:walberla}, {\it MUPHY}~\cite{bib:muphy}, and {\it Taxila} LBM~\cite{Coon2014}.  Interestingly, the first four
example implementations are able to handle solid objects suspended in fluids.
The first three are even able to combine this with multiple fluid components or
phases by using different LBMs. All five codes
demonstrated excellent scaling on hundreds of thousand CPUs available on state
of the art supercomputers.

{\it Ludwig} is a feature-rich implementation which was developed at the University of Edinburgh. It is based on the free-energy model~\cite{bib:swift-osborn-yeomans}. Recently, algorithms for interacting colloidal particles, following the method given in Sec.~\ref{Sec:particles}, have been added~\cite{bib:stratford-adhikari-pagonabarraga-desplat-cates-2005}. Similar in functionality, but based on the ternary Shan-Chen multicomponent model is {\it lb3d}, which was developed at University College London, University of Stuttgart and Eindhoven University of Technology. In addition to simulating solid objects suspended in multiphase flow, it has the ability to describe deformable particles using an immersed boundary algorithm~\cite{bib:jens-jansen-2011,bib:jens-frijters-2012a,KFGKH13}. Both codes are matrix based and follow the classical domain decomposition strategy utilizing the Message Passing Interface (MPI), where every CPU core is responsible for a cuboid chunk of the total simulation volume. A refactored version of {\it lb3d} with limited functionality that focuses mainly on multiphase fluid simulation functionalities, has recently been released under the LGPL~\cite{MTPlb3dSite}. {\it Taxila} LBM is an open source LBM code recently developed at LANL. It is based on PETSc, Portable, Extensible Toolkit for Scientific computation (http://www.mcs.anl.gov/petsc/). {\it Taxila} LBM
solves both single and multiphase fluid flows on regular lattices in both two and three dimensions,
and the multiphase module is also based on the Shan-Chen model, but includes
many advances including higher-order isotropy in the fluid-fluid interfacial terms,
an explicit forcing scheme, and multiple relaxation times. Very recently, a 3D fully parallel code based
on the color gradient model~\cite{Liu2012a}, {\it CFLBM}, has been developed jointly by UIUC and LANL. Like {\it Ludwig} and {\it LB3D}, the
{\it CFLBM} code is matrix based and follows the classical domain decomposition strategy utilizing
MPI. {\it CFLBM} has been run on LANL's Mustang and UIUC's Blue Waters using up to 32,768 cores, and exhibited nearly ideal scaling.
{\it MUPHY} was developed by scientists in Rome, at Harvard university and at NVidia and focuses on the simulation of blood flow in complex geometries using a single phase lattice Boltzmann solver. To model red blood cells, interacting point-like particles have been introduced. The code has demonstrated excellent performance on classical and GPU based supercomputing platforms and was among the finalists for several Gordon Bell prizes. In contrast to {\it Ludwig} and {\it lb3d} it uses indirect addressing in order to accommodate the complex geometrical structures observed in blood vessels in the most efficient way.
The code from the University of Erlangen, {\it walBerla}, combines a free surface multiphase lattice Boltzmann implementation with a solver for almost arbitrarily shaped solid objects. As an alternative to direct or indirect addressing techniques, it is based on a ``patch and block design'', where the simulation domain is divided into a hierarchical collection of sub-cuboids which are the building blocks for massively parallel simulations with load balancing~\cite{bib:walberla}.

\begin{figure}
  \centering
  \subfigure[]
  {\label{fig:speedup-nomd}\includegraphics[width=0.45\textwidth]{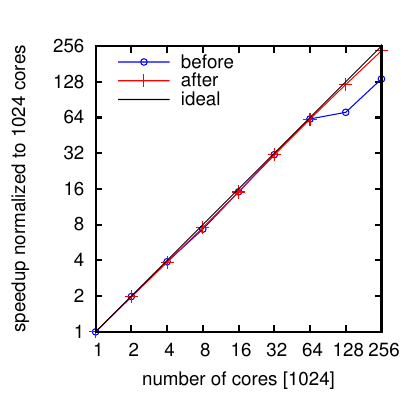}}
  \hfill
  \subfigure[]
  {\label{fig:speedup-md2}\includegraphics[width=0.45\textwidth]{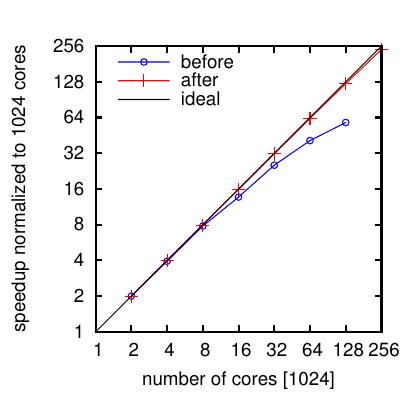}}
  \caption{Strong scaling of {\it lb3d} on a Blue Gene/P machine before and after optimizations. \subref{fig:speedup-nomd} relates to a system with only one fluid component. \subref{fig:speedup-md2} refers to a system with two fluid species and suspended particles  (from Ref.~\cite{bib:jens-groen-henrich-janoschek-coveney:2011}).}
\end{figure}

Some implementation details relevant to massively parallel simulations using the LBM are given with {\it lb3d} as an example.  The software is written in Fortran 90 and parallelized using MPI. To perform long-running simulations on massively parallel architectures requires parallel I/O strategies and checkpoint and restart facilities. {\it lb3d} uses the parallel HDF5 formats for I/O which has proven to be highly robust and performant on many supercomputing platforms worldwide.  Recently {\it lb3d} has
been shown to scale almost linearly on up to 262,144 cores on the European Blue Gene/P systems Jugene and Juqueen based at the J\"ulich Supercomputing Centre in Germany~\cite{bib:jens-groen-henrich-janoschek-coveney:2011}. However, to obtain such excellent scaling, some optimizations of the code were required. The importance of these implementation details is depicted by strong scaling measurements based on a system of $1\,024^2\times2\,048$ lattice sites carrying
only one fluid species (Fig.~\ref{fig:speedup-nomd}) and a similarly sized system containing two fluid species and $4\,112\,895$ uniformly distributed particles with a radius of five lattice units (Fig.~\ref{fig:speedup-md2}). Initially, LB3D showed only low efficiency in strong scaling beyond $65\,536$ cores of the BlueGene/P system. This problem could be related to a mismatch of the network topology of the domain decomposition in the code and the network actually employed for point-to-point communication. The Blue Gene/P provides direct links only between direct neighbors in a three-dimensional torus, so a mismatch can cause severe performance losses. Allowing \texttt{MPI} to reorder process ranks and manually choose a domain decomposition based on the known hardware topology, efficiency can be brought close to ideal. See Fig.~\ref{fig:speedup-nomd} for a comparison of the speedup before and after this optimization. Systems containing particles and two fluid species were known to slowly degrade in parallel efficiency when the number of cores was increased. This degradation was not visible for a pure lattice Boltzmann system, and could be attributed to a
non-parallelized loop over all particles in one of the subroutines implementing the coupling of the colloidal particles and the two fluids. Due to the low computational cost per iteration compared to the overall coupling costs for colloids and fluids, at smaller numbers of particles or CPU cores this part of the code was not recognized as a possible bottleneck.  A complete parallelization of the respective parts of the code produced a nearly ideal speedup up to $262\,144$ cores also for this system. Both strong scaling curves are depicted in Fig.~\ref{fig:speedup-md2}.

In order to improve the accuracy and numerical stability of \emph{lb3d} with respect to the application to the simulation of multiphase flows in porous media, recently a MRT collision model was integrated with the software. While the MRT collision algorithm is more complex than the BGK collision scheme and can cause significant performance loss when implemented naively, the increase in calculation cost can be dramatically reduced. We take advantage of two properties of the system to minimize the impact of the additional MRT operations on the code performance. First, the symmetry of the lattice allows prior calculation of the sum and difference of discrete velocities which are linear dependent, thus saving at least half of the calculation steps~\cite{PhysRevE.75.066705}. Second, the equilibrium stochastical moments can be expressed as functions of the conserved properties density and velocity, thus saving the transformation of the equilibrium distributions~\cite{bib:dhumieres-ginzburg-krafczyk-lallemand-luo}. As such, the performance penalty could be reduced below 17\%, which is close to the minimal additional cost reported
in~\cite{bib:dhumieres-ginzburg-krafczyk-lallemand-luo}. Since in multiphase systems the relative cost of the collision scheme is further reduced, the use of the MRT scheme has even less impact on the performance and for ternary amphiphilic simulations we find a performance penalty of only 5.8\%.

%
%

\subsection{Parallelization on general purpose GPU arrays}

The introduction of application programming interfaces such as CUDA, OpenCL, DirectCompute, and the addition of compute shaders in OpenGL, has enabled implementation of numerical methods on graphics processing hardware. When used for scientific computing, there are two primary advantages of a GPU over a CPU. Firstly, a GPU typically has a far greater number of cores. The current generation nVidia Tesla K20x has 2688 cores, while a high end Intel Xeon E-2690v2 has only 10. Second, GPUs also have a much higher memory bandwidth, with a theoretical maximum of 250 GB/s versus 59.7 GB/s for the Tesla and Xeon, respectively. It is therefore reasonable to expect that implementation of the LBM on a GPU architecture will yield significant performance improvements when compared with an equivalent multicore or cluster-based CPU implementation.

As with many CPU implementations, GPU parallelism requires the LBM domain to be decomposed into a number of equal sized blocks of lattice sites. The GPU hardware is partitioned into streaming multiprocessors (SM), each consisting of a number of cores. Each domain block is assigned to a single SM, where the lattice sites are assigned to a core and computed in parallel. The LBM computations are implemented as a \textit{kernel function} which is executed on the GPU.

The performance benefits of using GPUs with the LBM are well reported. Mawson \& Revell's implementation on a single Tesla GPU achieved a peak performance of 1036 million lattice updates per second (MLUPS)~\cite{Mawson2013}. Implementation of the method by Obrecht \textit{et. al.} on a GPU cluster, with an older generation of GPUs, yielded speeds in excess of 8,000 MLUPS~\cite{Obrecht2013259}. The work detailed by these authors reveals that writing an efficient kernel function is, however, non-trivial. Indeed, a number of issues, such as branching code, memory access, and memory consumption, must be considered when writing an efficient LBM kernel.

Branching code refers to the use of conditional statements to direct the logic of an algorithm. When a conditional statement (e.g. if statement) is executed on a CPU only the valid branch of code will be computed. This is not the case with a GPU. GPU's are designed to be \textit{Same Instruction Multiple Data} machines. This means that all cores in an SM must execute the same code which leads to two possible outcomes. In the case that all cores evaluate the statement and require the same branch of the conditional statement, only one branch is actually computed. If some cores require the first branch of the statement, and the rest require the second branch, then all cores execute \textit{both} branches of the conditional statement. In the latter case, the redundant branch is computed as a \textit{null pointer operation}.

There are a two situations where this branching problem is particularly relevant to the implementation of a general LBM code, namely the collision operator and boundary conditions. It is common for a general code to implement a variety of collision operators. However, the use of multiple collision operators within a single model is uncommon. In this instance it is acceptable for the collision operator selection logic to appear within the kernel. As in this case, all threads will only execute a single branch. The selection of boundary conditions at a node is also done through conditional logic. The naive approach to avoiding code branching in this scenario is to simply use two separate kernels for boundary and regular lattice sites. However, this approach requires the development of more code, and the spatial locality of lattice sites is not preserved in memory.

As was mentioned in Sec. \ref{sec:multicore}, many LBM implementations use two data structures to store particle distribution functions. This is not optimal when using GPU hardware, as even the most recent Tesla GPUs are limited to 6GB of memory, which corresponds to approximately  34 million lattice sites when using a D3Q19 lattice. Fortunately, a number of approaches exist to remove the dual-lattice requirement. These include the \textit{Compressed Grid} or \textit{Shift} algorithm proposed by Pohl \textit{et al.}~\cite{Pohl2003}, the \textit{Swap} algorithm proposed by Latt~\cite{Latt2007a}, and the \textit{AA Pattern} algorithm proposed by Bailey \textit{et al.}~\cite{Bailey2009}. These approaches have been reviewed by Wittmann \textit{et al.} who found that of the algorithms available, the AA Pattern algorithm proved to be the most beneficial both in terms of memory consumption and computational efficiency~\cite{Wittmann2011}.

In order to achieve the maximum theoretical memory bandwidth of a GPU, memory access must be \textit{coalesced}. An access pattern which is coalesced is one where, for double precision, accesses fit into segments of 128 bytes resulting in threads that read data from the same segment of the array. Various authors have presented methods to mitigate the impact of uncoalesced memory access. Toelke \textit{et al.} presented a method where the streaming operation was split into two stages for a $D2Q9$ lattice, where variables are streamed first in the X-direction, and subsequently in the Y-direction~\cite{Tolke2008}. Rinaldi \textit{et al.} noted that uncoalesced memory reads are faster than uncoalesced writes, so they carried out propagation of the particle distribution functions in the reading step~\cite{Rinaldi2012}. Streaming on read is a straightforward approach that mitigates the effect of coalesced access at no extra cost. However, recent work by Mawson \& Revell has shown that techniques like the one proposed by Toelke \textit{et al.} require extra processor registers, limiting the number of lattice sites which may be computed in parallel. Mawson \& Revell found that, with the current generation of nVidia Tesla chips, the bandwidth reduction due to uncoalesced streaming in the LBM is at most 5\%~\cite{Mawson2013}.

Finally, a multi-GPU configuration can be useful when either the memory consumption of a model exceeds the available memory on a single GPU, or more computational power is required. For a code to exploit multiple GPUs another level of domain decomposition must be added in which each GPU is used to compute a subdomain. To do this, the storage strategy for lattice data must account for communication between GPUs. The solution to managing this process is found in the way in which data is marshalled between nodes in a cluster. In this case, the CPU represents a master node, the GPUs then become the slave nodes. Where the bandwidth of the PCIe x16 ports used to connect the GPU is of a similar order of magnitude to that fourteen data rate InfiniBand connection at 15.4GB/s. Though unlike an InfiniBand connection, it is not currently possible for slave nodes to access the memory of other slave nodes directly. The best strategy for this is to include \textit{ghost nodes} so that race conditions are avoided during the streaming of particle distribution functions across the subdomain boundary. The master node, or CPU, is then used to manage data transfer between devices.

\section{Applications}
\label{sec:app}

In this section the ability of the discussed multiphase models to capture the relevant physics of multiphase flow problems in porous media is demonstrated. This is undertaken using baseline tests of two-dimensional flow in synthetic porous media.

\subsection{The color gradient model}

\begin{figure}
\includegraphics[width=0.75\textwidth]{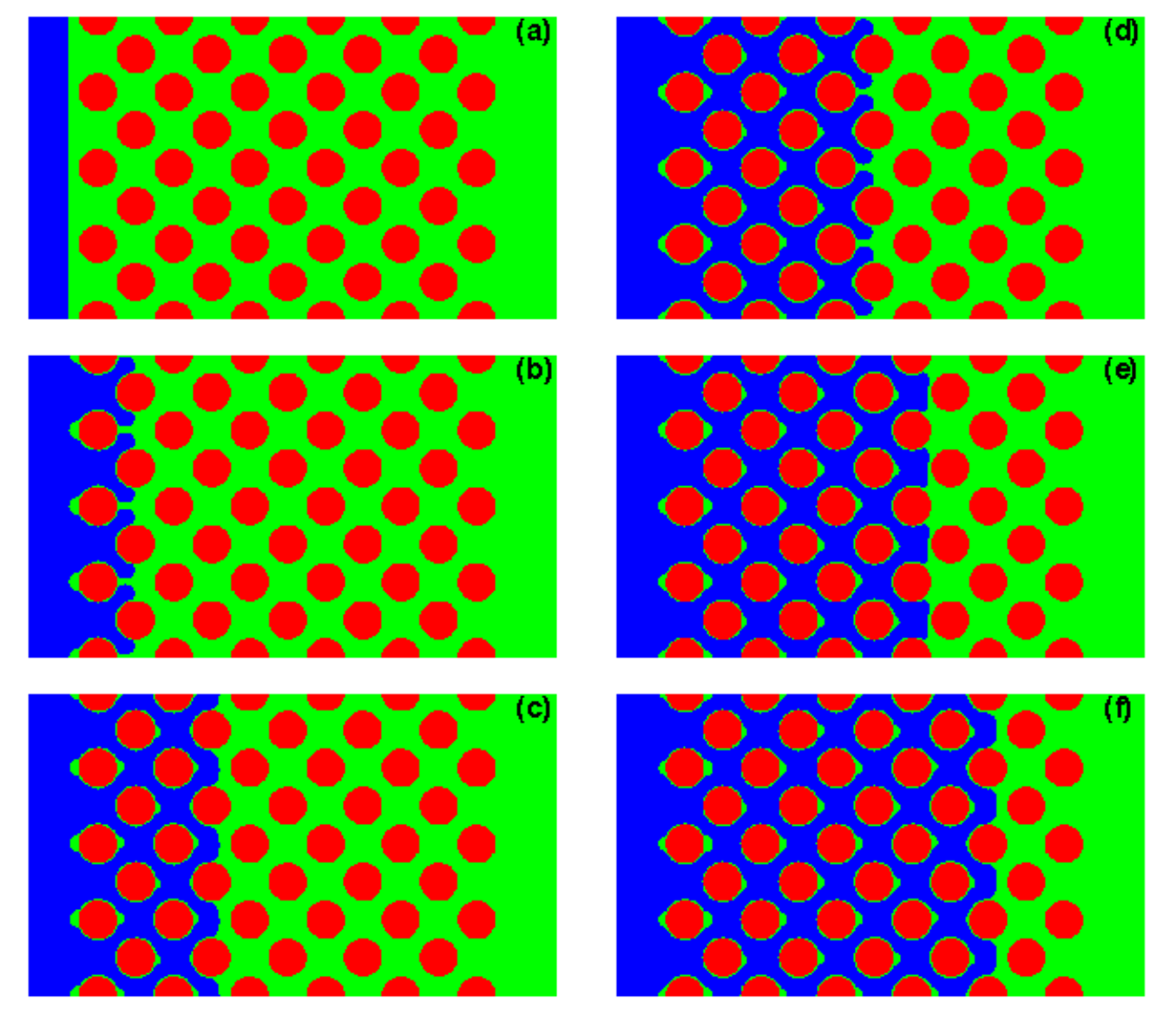}
\centering \caption{(Color online) Time evolution of non-wetting fluid displacing
the wetting fluid for $Ca=5\times 10^{-3}$ and $M=\frac{1}{10}$ at
timesteps: (a) 0;~(b)16000;~(c)34000;~(d)48000;~(e)60000;
and~(f)74000.} \label{fig2}
\end{figure}

\begin{figure}
\includegraphics[width=0.75\textwidth]{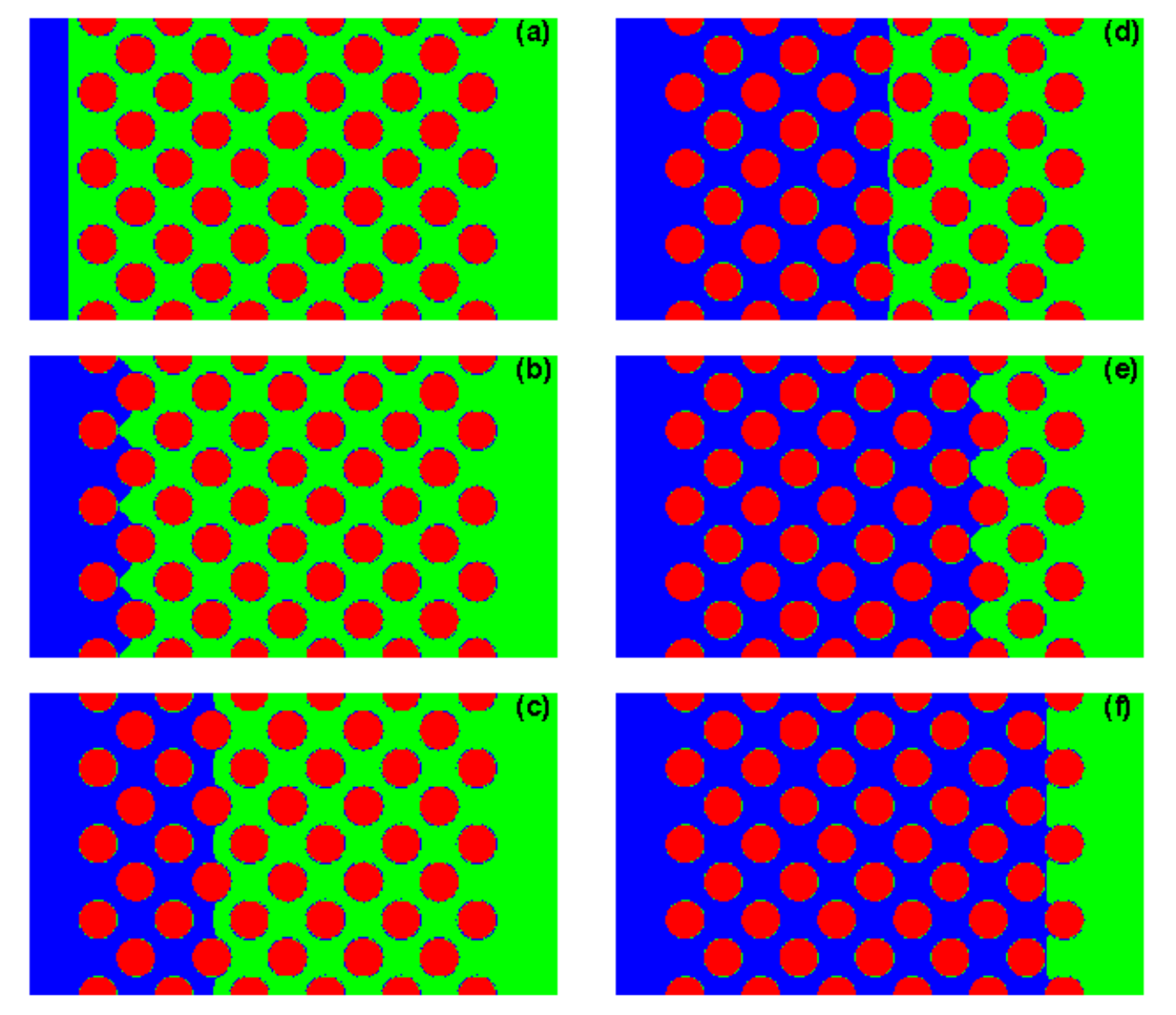}
\centering \caption{(Color online) Time evolution of wetting fluid displacing the
non-wetting fluid for $Ca=5\times 10^{-3}$ and $M=4$ at timesteps: (a)
0;~(b)8000;~(c)18000;~(d)26000;~(e)38000; and~(f)46000.}
\label{fig3}
\end{figure}

In the current example,
no-slip boundary conditions at solid walls are implemented by a simple bounce-back rule~\cite{Latva-Kokko2005a}.
The wettability of the solid walls can be imposed by assuming that the solid wall is a mixture
of two fluids, thus having a certain value of the phase
field~\cite{Latva-Kokko2005a,Liu2011}. The interfacial force term in Eq.~(\ref{eq_liu}) becomes dependent on the
properties of the neighboring solid lattice sites, resulting in a
special case of the wetting boundary condition.
The
assigned value of the phase field at sites neighboring the wall sites can be used to
modify the static contact angle of the interface.
Figs.~\ref{fig2} and \ref{fig3} give the time evolution of interface at $Ca=0.005$
for drainage process with $M=\frac{1}{10}$ and imbibition process
with $M=4$ in a 2D pore network, consisting of a uniform
distribution of circular grains. Here, the capillary number ($Ca$)
relates viscous to capillary forces and is defined as
$Ca=u_{in}\eta_{in}/\sigma$, where $u_{in}$ and $\eta_{in}$ are
the mean velocity and dynamic viscosity of the displacing fluid at the
inlet, respectively. The viscosity ratio, $M$, is defined as the
ratio of non-wetting fluid viscosity to the wetting fluid
viscosity: $M=\eta_{nw}/\eta_{w}$. In the drainage process, the non-wetting
fluid advances in a piston-like manner in the pore throats. It can be clearly
observed that there is one small blob of defending fluid trapped near the rear
stagnant point for each solid grain. Similar trapped blobs are also found at
the front stagnant point for the first column of grains. These trapped blobs of
defending fluid are attributed to low flow velocity and high pressure at the
front and rear stagnant points, so that the wetting fluid cannot be completely
drained out before the advancing interfaces of non-wetting fluid coalesce or
touch the surface of solid grains. However, the defending fluid is completely
drained out in the imbibition process, where the interface advances in a more
stable manner, and the solid surface favors the invading fluid but repels the
defending fluid.

\subsection{Inter-particle potential model}
The inter-particle potential model was used extensively for various multiphase
flow problems, see
Refs.~\cite{Martys1996,Schmieschek2011,Pan2004,Sukop2004,Li2005,Huang2009a,Porter2009,Parmigiani2011,Dong2011,Ghassemi2011,Mukherjee2011,Middleton2012,Dou2013,bib:jens-venturoli-coveney:2004,bib:jens-harvey-chin-venturoli-coveney:2005},
because of its simplicity and easy implementation. Pan et al.~\cite{Pan2004}
used the MCMP inter-particle potential model to simulate two-phase flow in a
porous medium comprised of a synthetic packing with a relatively uniform
distribution of spheres. They achieved good agreement between the measured
hysteretic capillary pressure saturation relations and the lattice Boltzmann simulations when
comparing entry pressure, displacement slopes, irreducible saturation, and
residual entrapment.
The hysteresis was also
found by Sukop and Or~\cite{Sukop2004} who adopted the SCMP inter-particle
potential model to simulate the liquid-vapor distributions in a porous medium
based on two-dimensional imagery of a real soil. Porter et
al.~\cite{Porter2009} further emphasized the importance of the
wetting-nonwetting interfacial area. They adopted the MCMP inter-particle
potential model to study the hysteresis in the relationship between capillary
pressure, saturation, and interfacial areas in a three-dimensional glass bead
porous medium obtained by computed micro-tomographic (CMT) image data.

The inter-particle potential model has also been used to determine relative
permeabilities~\cite{Li2005,Ghassemi2011,Huang2009a,Dou2013}. Effects of
capillary number, wettability, and viscosity ratio, as well as the porous
structures on the relative permeability were investigated in detail. However,
the original inter-particle potential model suffers from some limitations,
including large spurious velocities in the vicinity of the fluid-fluid
interface, viscosity-dependent equilibrium density and interfacial tension, and
numerical instability for large viscosity or density ratios. In the SCMP model,
the kinematic viscosity is fixed, and the density ratio is limited to the
order of 10. In the MCMP model, the viscosity and density ratios are typically
restricted to no more than 5 and 3, respectively.
\begin{figure}
  \subfigure[]{ \label{fig:scImb1}
    \includegraphics[width=0.48\textwidth]{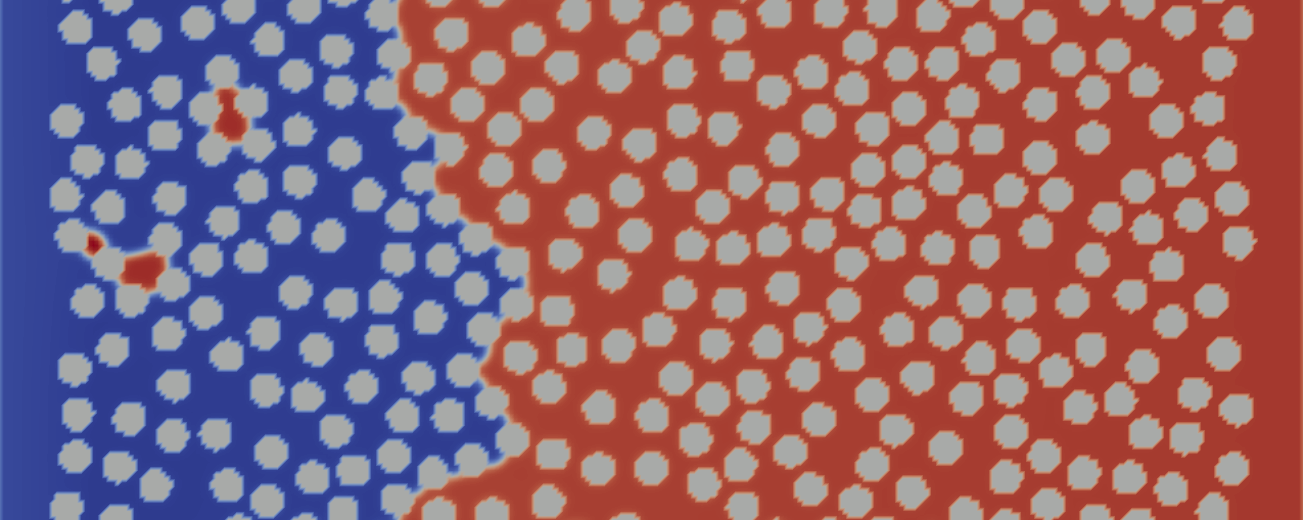}
    \includegraphics[width=0.48\textwidth]{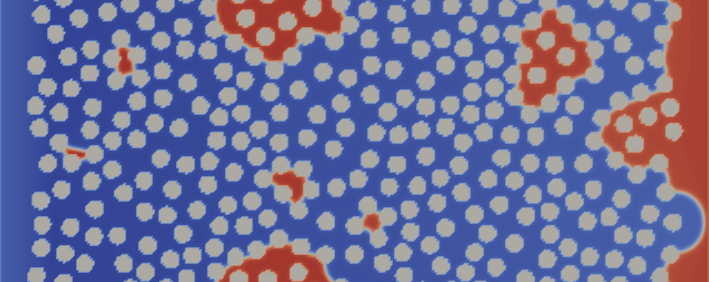}
  }
  \subfigure[]{ \label{fig:scImb2}
    \includegraphics[width=0.48\textwidth]{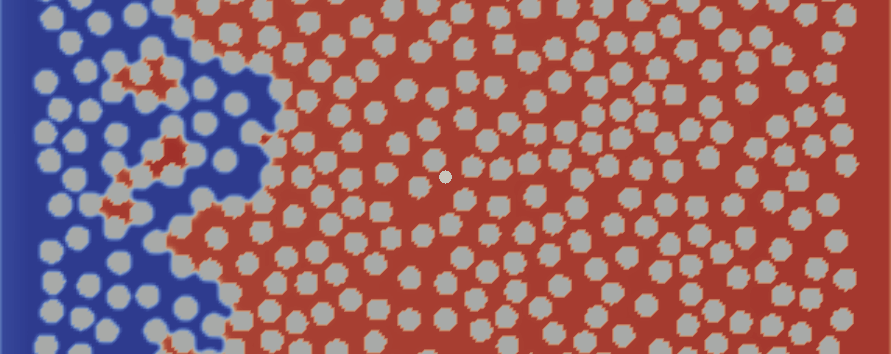}
    \includegraphics[width=0.48\textwidth]{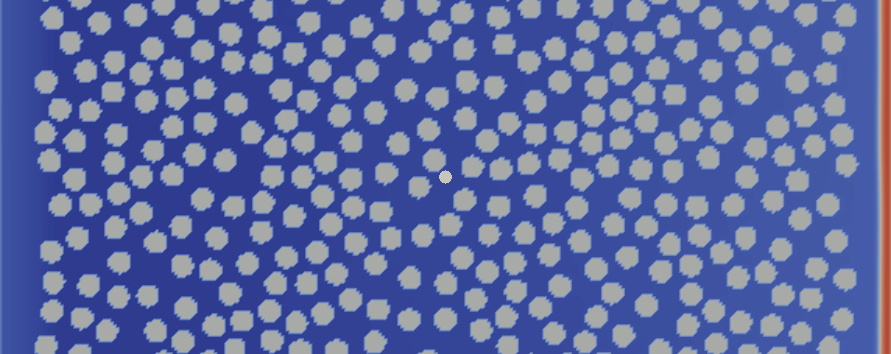}
  }
  \subfigure[]{ \label{fig:scImb3}
    \includegraphics[width=0.48\textwidth]{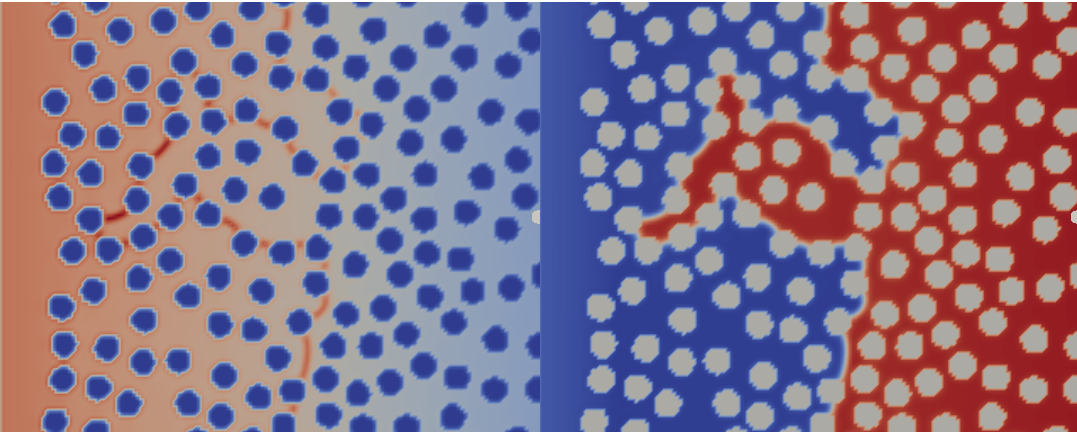}
    \includegraphics[width=0.48\textwidth]{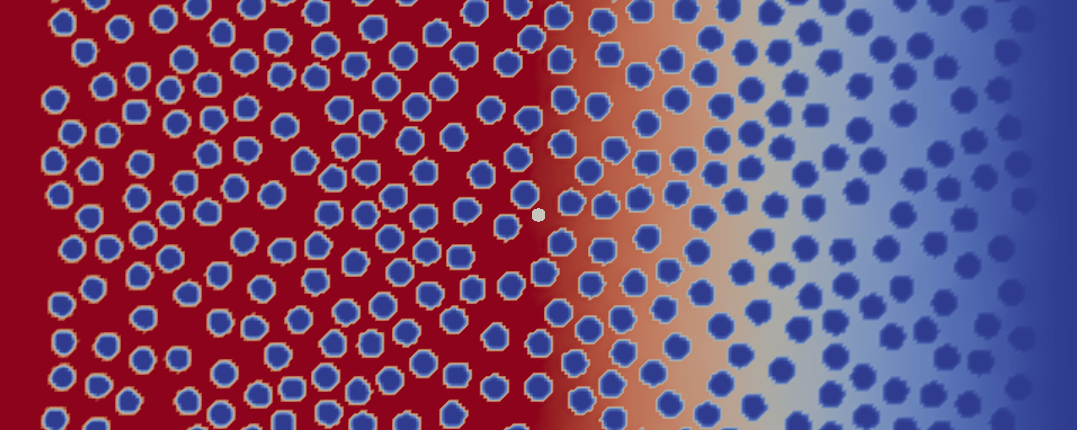}
  }
\caption{Snapshots of fluid density fields after imbibition simulation using an
inter-particle potential binary (ternary) fluid model in a pseudo-2d porous
medium, varying different parameters. \ref{fig:scImb1}: Comparison of stable
density distributions of the displaced (oil) component after 500,000 timesteps
varying an applied body force (pressure gradient) in lattice units between $F =
2\cdot 10^{-4}$ and $F = 4\cdot 10^{-4}$ in a neutrally wetting system.
\ref{fig:scImb2}: Comparison of stable density distributions of the displaced
(oil) component after 500,000 timesteps varying the contact angle of the
displaced component between $\Theta = 17^{\circ}$ and $\Theta = 163^{\circ}$,
applying a constant body force of $F = 4\cdot 10^{-4}$.
\ref{fig:scImb3}:Illustration of the surfactant component density field after
100,000 timesteps. Left: Surfactant component density (left) vs. displaced (oil)
component density (right) injecting a surfactant concentration
$c_{\text{S}}=0.2$ together with the displacing fluid into a neutrally wetting
system, applying a constant body force of $F = 4\cdot 10^{-4}$. Right:
Surfactant component density field injecting a surfactant concentration
$c_{\text{S}}=0.5$ together with the displacing fluid into a neutrally wetting
system, applying a constant body force of $F = 4\cdot 10^{-4}$.\label{fig:scImb}}
\end{figure}

Fig.~\ref{fig:scImb} illustrates simulations of imbibition into a pseudo-2D
porous medium using a ternary fluid mixture model as described in
Section~\ref{sec:mult}, equations
(\ref{eq:sccompsurfforce}-\ref{eq:scsurfsurfforce}). The qualitative effect of
variation of different system parameters on the stable configuration is being
investigated. The system consists of a 512$\times$1280 lattice with
randomly placed cylinders, which assures a minimum required resolution of the smallest pores. Re-coloring boundary conditions are applied at the
inlet and outlet so that fluid of one component crossing the periodic boundary
is added to the second component when appearing on the other side of
the system.  The surfactant follows standard periodic boundary conditions. The
coupling parameters of the inter-particle potential model are kept fixed at
$G_c=0.1$, $G_{k,s}=-0.006$, and $G_{s,s}=-0.003$.

Fig.~\ref{fig:scImb1} shows a comparison of the stable density
distributions of the displaced (oil) component after 500,000 timesteps. Here,
the applied body force, which is directly proportional to the pressure
gradient, is in lattice units varied between $F = 2\cdot 10^{-4}$ (left)
and $F = 4\cdot 10^{-4}$ (right). No surfactant is present and a contact angle
of $\Theta=90^{\circ}$ corresponding to neutral wetting is applied. Between
the considered values of forcing a transition from halted to complete filling
in the stable state of the system by the injected (water) component is
observed.

Again, Fig.~\ref{fig:scImb2} shows a comparison of the stable density
distributions of the displaced (oil) component after 500,000 timesteps. Here,
however, the driving body force is kept fixed at $F = 4\cdot 10^{-4}$ and the contact
angle of the displaced (oil) component is varied between $\Theta = 17^{\circ}$
and $\Theta = 163^{\circ}$. The strongly wetting case as depicted on the left
reverts the effect of stronger pressure and the system again becomes stable in
a partially filled state. As to be expected for the strongly dewetting case
shown on the right, the displaced (oil) component is completely flushed out of
the system.

Fig.~\ref{fig:scImb3} contains simulation snapshots after 100,000 timesteps.
The driving force and contact angle are kept constant at $F = 4\cdot 10^{-4}$,
and $\Theta=90^{\circ}$, respectively. A surfactant component is being added
to the invading fluid (water) component. On the left hand side for injecting a
surfactant concentration $c_{\text{S}}=0.2$, concurrent density fields of the
surfactant component (left) and the displaced (oil) component are plotted side
by side. The surfactant is agglomerating at the interface as denoted by the
dark red spots. For this concentration, a relatively sharp interface is
conserved. On the contrary, for a concentration of the surfactant component in
the invading fluid of $c_{\text{S}}=0.5$, a transition to a diffusive regime can be observed on the right hand side.
It is noteworthy that the interfacial width in the diffusive regime is of the
order of tens or even hundreds of lattice sites. In real life conditions, this should still
correspond to length scales of several nanometres or tens of nanometres only,
which depicts a limitation of diffuse interface models for porous media flows. 

These examples clearly demonstrate the ability of the amphiphilic model to qualitatively study the effect of surfactants on imbibition in porous media, which is of relevance for enhanced oil recovery applications.

\subsection{Free-energy model}

\begin{figure}
\includegraphics[width=0.7\textwidth]{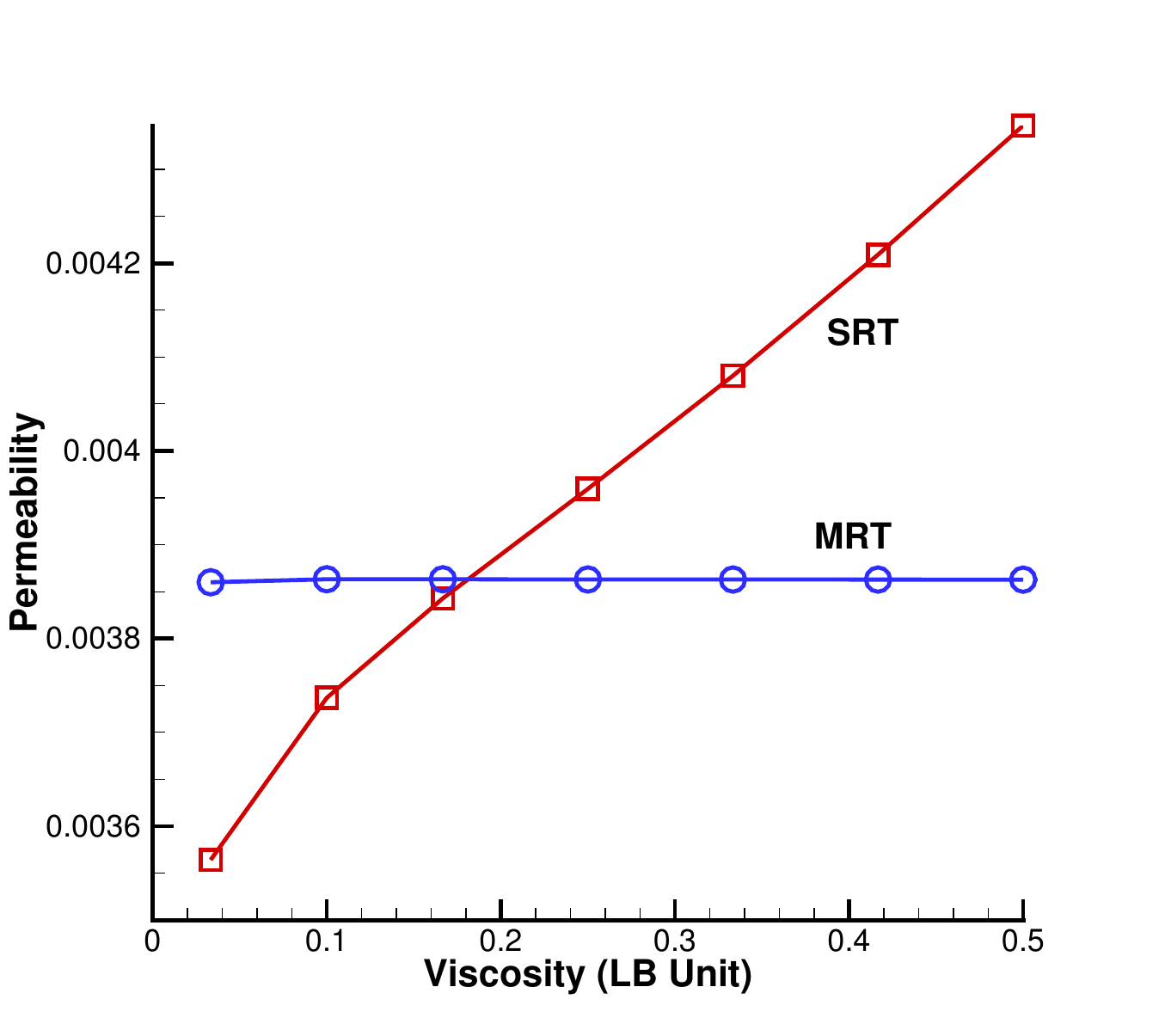}
\centering \caption{(Color online) Comparison of the permeability calculated based on MRT and SRT models with various viscosities for single-phase flow through a periodic body-centered cubic array of spheres, whose porosity is $\varepsilon=0.394$.}
\label{fig10}
\end{figure}

\begin{figure}
\includegraphics[width=0.8\textwidth]{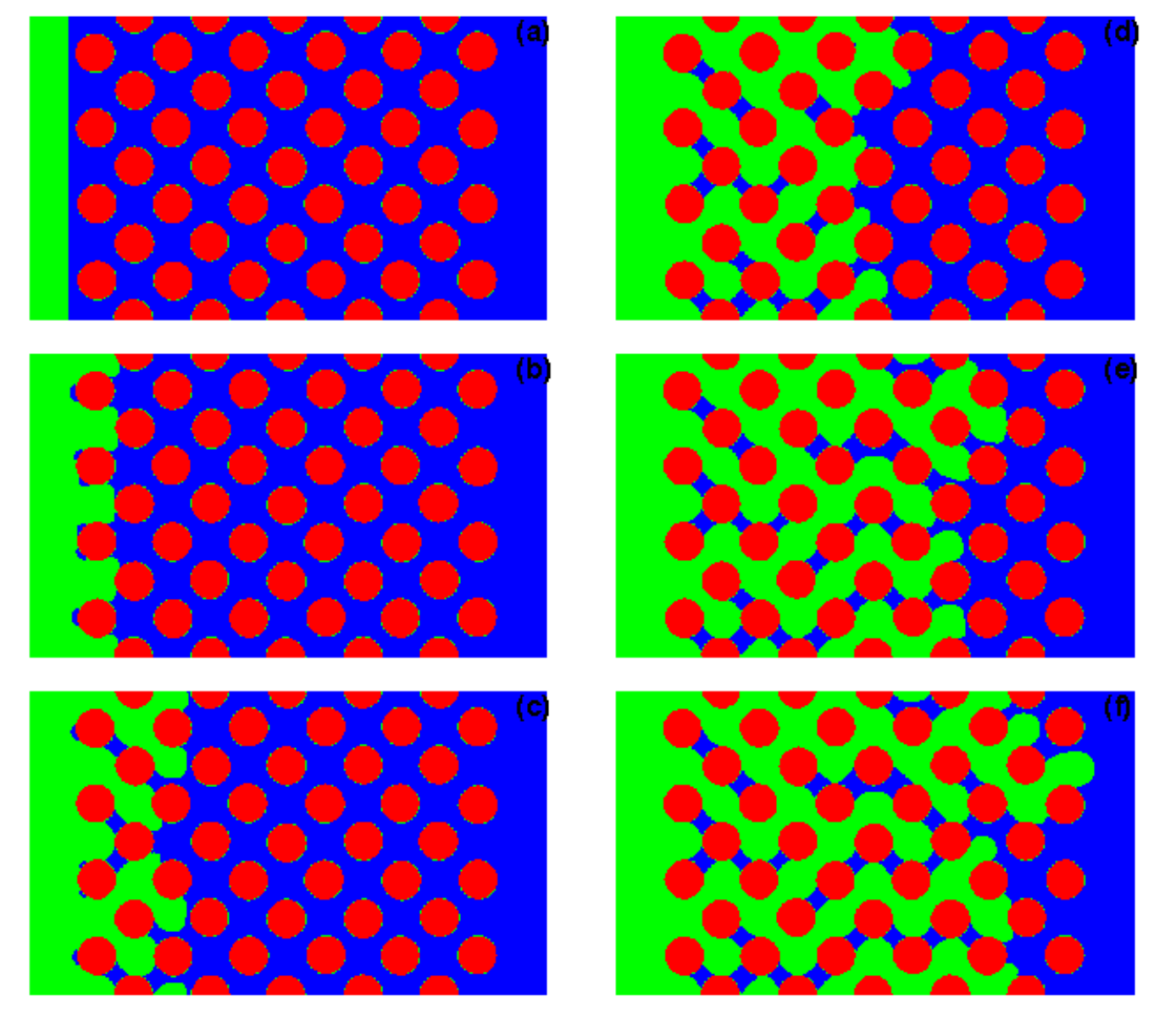}
\centering \caption{(Color online) Time evolution of less viscous non-wetting fluid displacing more viscous wetting fluid for $Ca=5\times 10^{-4}$, $M = 1/10$ and $\theta=135^{\circ}$ at timesteps (a)
0, (b) 200000, (c) 400000, (d) 800000, (e) 1100000, and (f) 1300000 using MRT free energy model of potential form.}
\label{fig11}
\end{figure}

It has been widely demonstrated that with the bounce-back boundary condition at
the solid walls, the SRT LBM produces viscosity-dependent
permeability in porous media, while the viscosity-independent solution can be
produced by MRT~\cite{Pan2006,Hao2009a}. This can be clearly seen in
Fig.~\ref{fig10}, which plots the permeability as a function of viscosity for
the single-phase flow through a body-centered cubic array of spheres. To produce viscosity-independent permeability, we implement the free-energy model of potential form using MRT with two independent relaxation times (i.e. the two-relaxation-time algorithm~\cite{Pan2006,Hao2009a}). This model is applied to simulate the less viscous non-wetting fluid displacing the wetting
fluid in a pore network with slightly irregular distribution of cylinder center
(which is obtained by adding randomly small perturbation to the regularly
distributed position).  Fig.~\ref{fig11} gives the displacement processes for
the periodic boundary conditions at the upper and lower boundaries. Similar to
the observations in the color gradient model, we can clearly see some small blobs
trapped near the front sides for the first column of solid grains. Also, the
trapped small blobs can dissolve very quickly as the simulations progress.
Actually, the dissolution of small droplets/bubbles is a typical phenomenon in many diffuse-interface models (e.g. the phase-field or free-energy model). The droplet dissolution is attributed to two factors. The first is that a multiphase system is
always evolving towards the direction of decreasing free energy in the
free-energy model, and the system with the droplets completely
dissolved has a lower free energy than the one with two-phase coexistence, so
small droplets are prone to dissolve~\cite{Sman2008}. The second is that the Cahn-Hilliard
equation can conserve the total mass of the system but cannot conserve the mass
for each component/fluid. Several methods have been proposed for reducing the
rate of dissolution in some simple systems, but more efforts are still required
to obtain physically meaningful numerical results in a large and complex porous
media, where the slim fingers may be dynamically evolving and sometimes unstable.

\subsection{Mean-field theory model}

\begin{figure}
\includegraphics[width=0.8\textwidth]{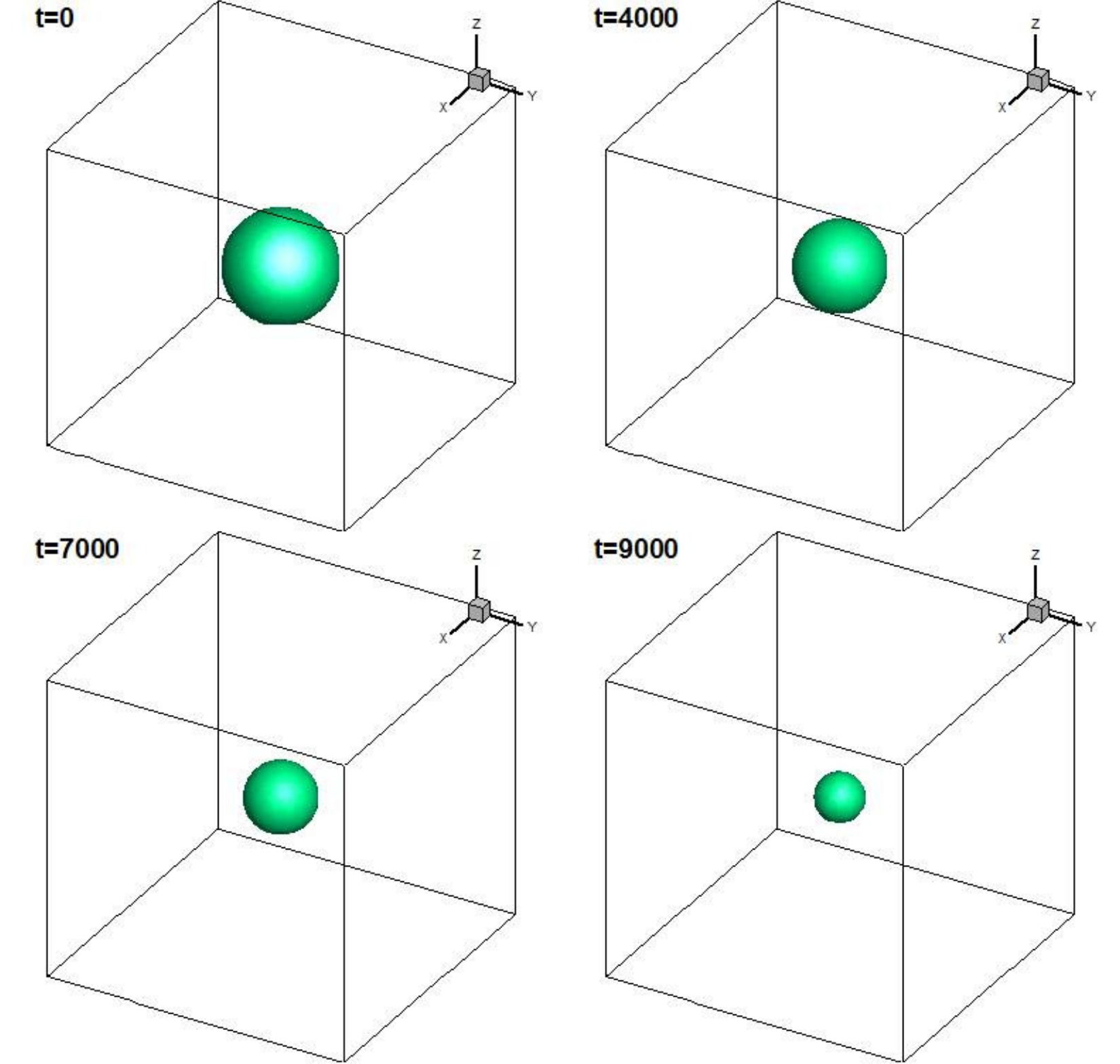}
\centering \caption{(Color online) Time evolution of a liquid droplet with radius $R = 10$ in a stationary gas phase.
The size of computational domain is $60\times 60 \times60$. The densities of liquid and gas
are 1 and 0.25, and the relaxation times for both fluids are taken 0.53. Other parameters can be found in Ref.~\cite{Premnath2007}.}
\label{fig12}
\end{figure}

The mean-field theory model has been implemented by Premnath and
Abraham~\cite{Premnath2007} with the MRT algorithm in order to achieve better
numerical stability. Using this MRT model, we also simulated a 3D stationary
bubble in a liquid domain with periodic boundary conditions applied at all the
boundaries. When using the same parameters as given in~\cite{Premnath2007}, we
found that good results can be obtained without bubble dissolution.  However,
as the bubble size is decreased, it can be observed that the bubble can quickly
dissolve, which is shown in Fig.~\ref{fig12}. A fast dissolution is disastrous
for obtaining  reliable simulation results. Fakhari and
Rahimian~\cite{Fakhari2010a} also noticed the dissolution problem in their
two-dimensional axisymmetric simulations, and they suggested to take $a =
12.75RT$, which can effectively reduce diffusion of different phases into each
other. However, we found that this improvement is not very effective in 3D
stationary bubble tests.

\subsection{Stabilized diffuse-interface model}
\begin{figure}
\includegraphics[width=0.9\textwidth]{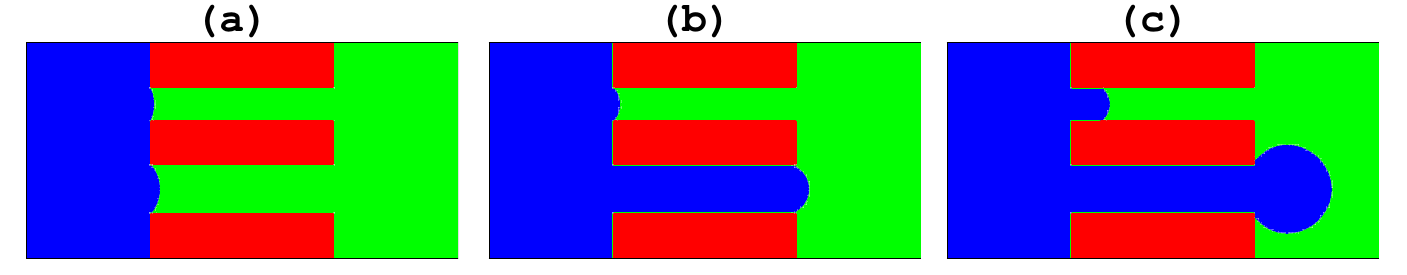}
\centering
\caption{(Color online) Injection of a non-wetting gas into two
  parallel capillary tubes with the pressure difference $\Delta p$ of
  (a)~$4\times 10^{-5}$, (b)~$6\times 10^{-5}$, and (c)~$8\times
  10^{-5}$. The capillary pressure is $P_{c_1}=7.1\times 10^{-5}$ for
  the upper tube and $P_{c_2}=4.7\times 10^{-5}$ for the lower tube
  (reproduced from Ref.~\cite{Liu2013}).}
\label{fig13}
\end{figure}

\begin{figure}
\includegraphics[width=0.75\textwidth]{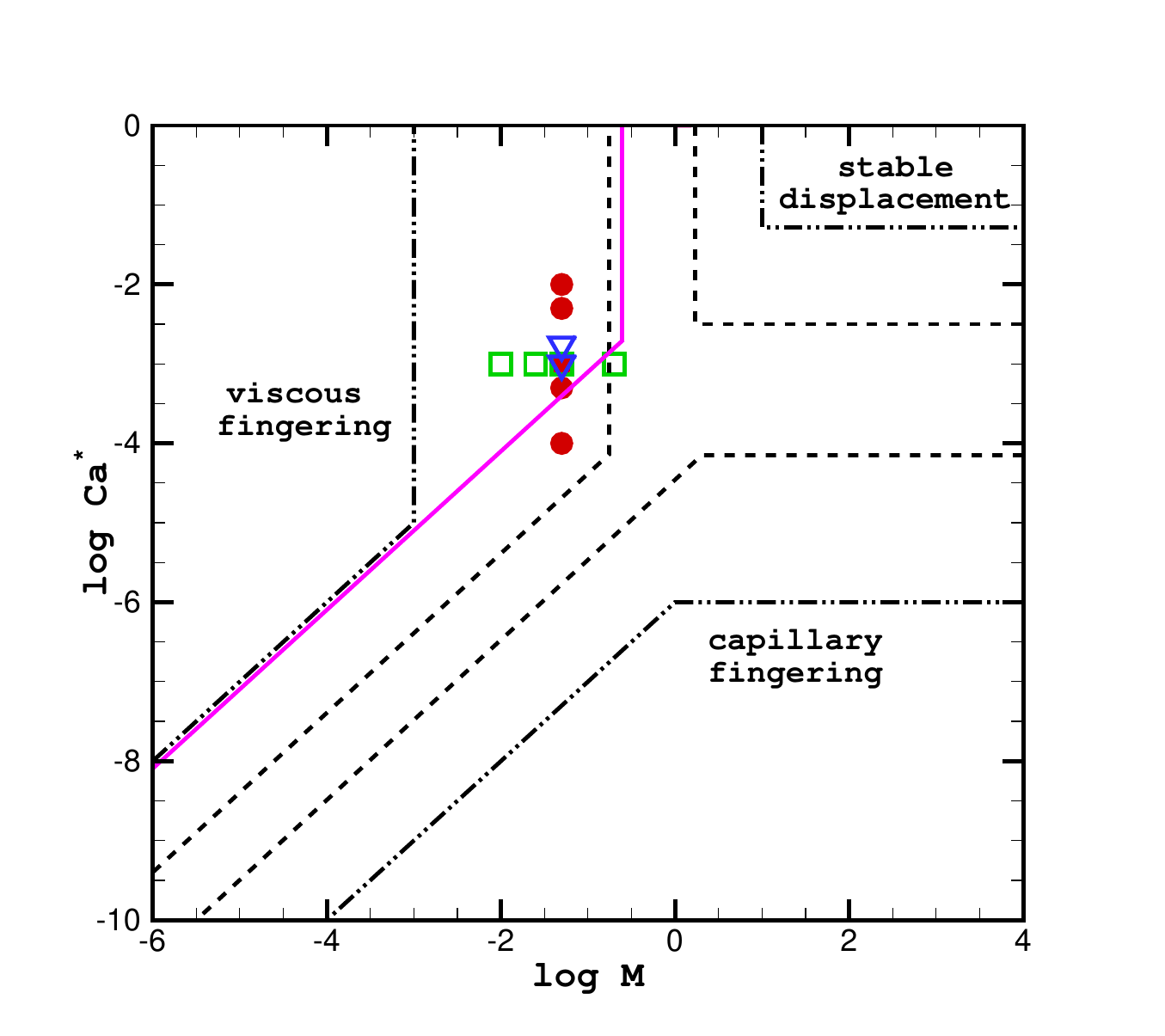}
\centering
\caption{(Color online) $\log Ca^*-\log M$ phase diagram indicating
  the fluid displacement patterns and the locations of the numerical
  simulations given by Liu et al.~\cite{Liu2013} (represented by
  discrete symbols) for drainage displacement, where the capillary
  number $Ca^*$ is defined by $Ca^*=\frac{u_{in}\eta_{in}}{\sigma
    \cos(\theta)}$. The stability zones bounded by black
  dash-dot-dotted, black dashed, and pink solid lines, are obtained by
  Lenormand et al.~\cite{Lenormand1988}, Zhang et
  al.\cite{Zhang2011a}, and the simulations of Liu et al.,
  respectively (reproduced from Ref.~\cite{Liu2013}).}
\label{fig14}
\end{figure}

Here, we demonstrate that the stabilized diffuse-interface model is
most suitable to simulate flow problems with high density ratio.
No-slip boundary conditions are applied at solid walls using the
bounce-back scheme~\cite{Lee2010}. For a straight solid wall, the
method of Lee and Liu~\cite{Lee2010} can be employed to impose the
wetting boundary condition. Recently, a wetting boundary treatment was
proposed for concave and convex corners, which can be extended to more
complicated geometries with curved boundaries represented by a
staircase approximation~\cite{Liu2013}. With the proposed wetting
boundary treatment, Liu et al.~\cite{Liu2013} have simulated the
injection of a non-wetting gas through two parallel capillary tubes
(the widths of the upper and lower capillaries are $r_1$ and $r_2$,
and $r_1<r_2$, leading to the capillary pressure $p_{c1}>p_{c2}$.) at
several different $\Delta p$, where $\Delta p$ is the pressure
difference between the inlet and outlet.  As expected, the findings
were that when $\Delta p$ is smaller than $p_{c2}$ (Fig.~\ref{fig13}(a)), the invading gas cannot enter both capillary tubes, when $\Delta p$ is
between $p_{c2}$ and $p_{c1}$ (Fig.~\ref{fig13}(b)), the gas only
flows into the large capillary tube, and when the pressure difference is
increased to $\Delta p>p_{c1}$ (Fig.~\ref{fig13}(c)), the gas flows
into both capillary tubes, but the displacement is much faster in the
large capillary tube. This displacement behavior is consistent with
the principle of pore-network simulators~\cite{Lenormand1988}, which
suggests that this HDR model is able to capture capillary effects and
reproduce correct displacement behavior. The stabilized
diffuse-interface model was also used to simulate gas displacement of
liquid in a homogenous two-dimensional pore network consisting of
uniformly spaced square obstructions. The effect of capillary number,
viscosity ratio, surface wettability, and Bond number was studied
systematically. Similar to previous experimental
observations~\cite{Lenormand1988}, three different regimes, namely
stable displacement, capillary fingering, and viscous fingering, were
identified in the drainage displacement, and all of them are strongly
dependent upon the capillary number, viscosity ratio, and Bond
number. The simulation results shown in the two-dimensional phase
diagram (see Fig.~\ref{fig14}) denote that the viscous fingering regime
covers a region markedly different from those obtained in previous
numerical and experimental studies~\cite{Lenormand1988,Zhang2011a}.
The difference is because the boundaries of the regimes in the phase
diagram are strongly dependent on the configuration of the pore
network, and also upon 3D effects, which are neglected in 2D
simulations but can play a non-trivial role for determining the
displacement behavior in micromodel laboratory experiments.

\begin{figure}
  \subfigure[tracer]{ \label{fig:trj1}
    \includegraphics[width=0.48\textwidth]{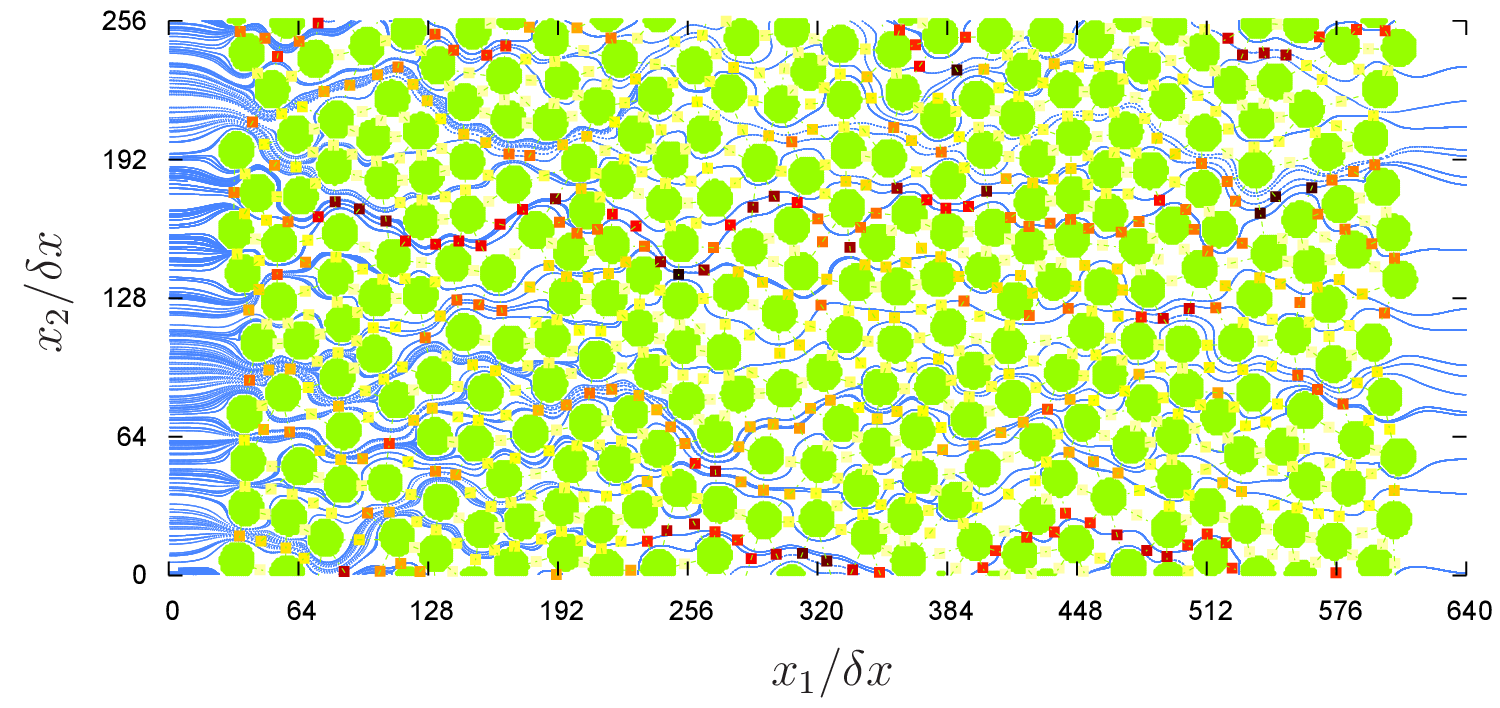}
  }
  \subfigure[$r_{\rm p}=1.6$]{ \label{fig:trj3}
    \includegraphics[width=0.48\textwidth]{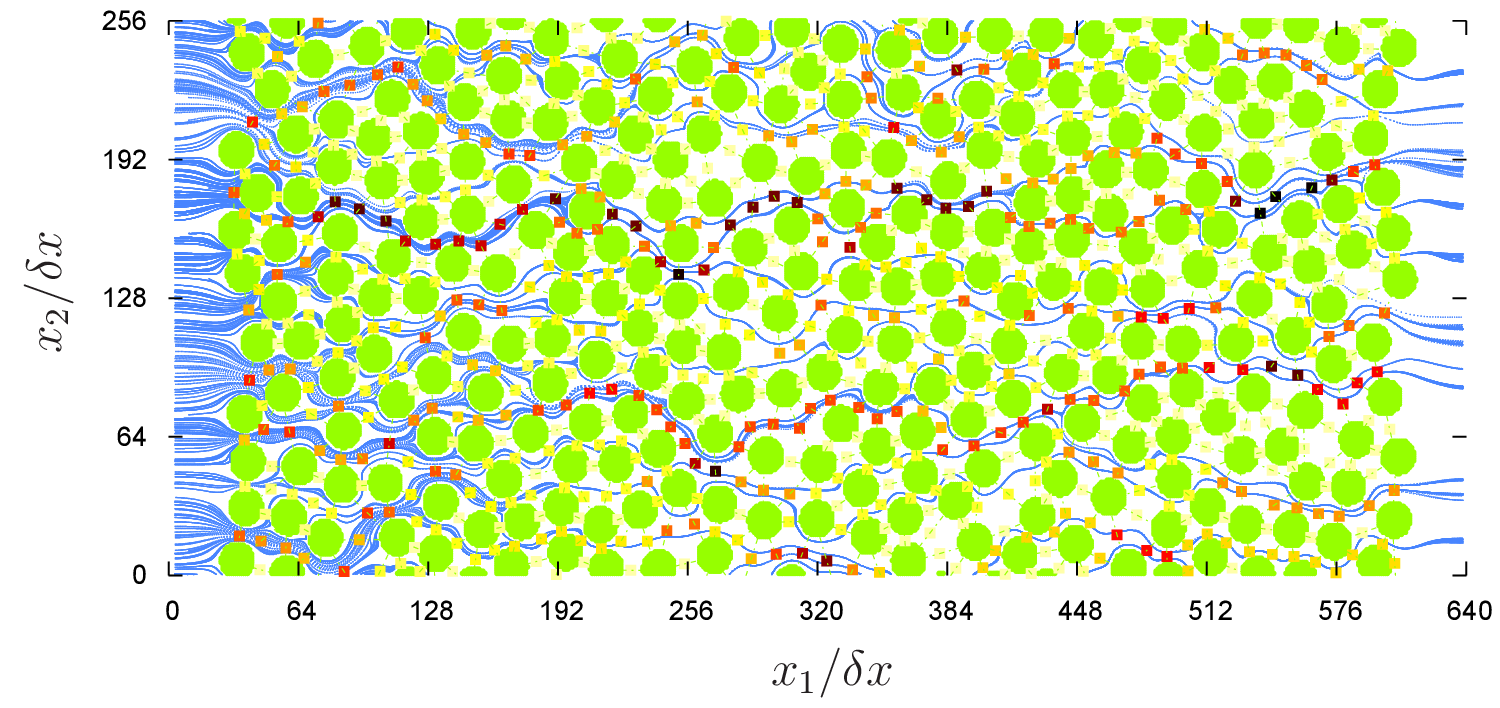}
  }
  \subfigure[$r_{\rm p}=2.1$]{ \label{fig:trj2}
    \includegraphics[width=0.48\textwidth]{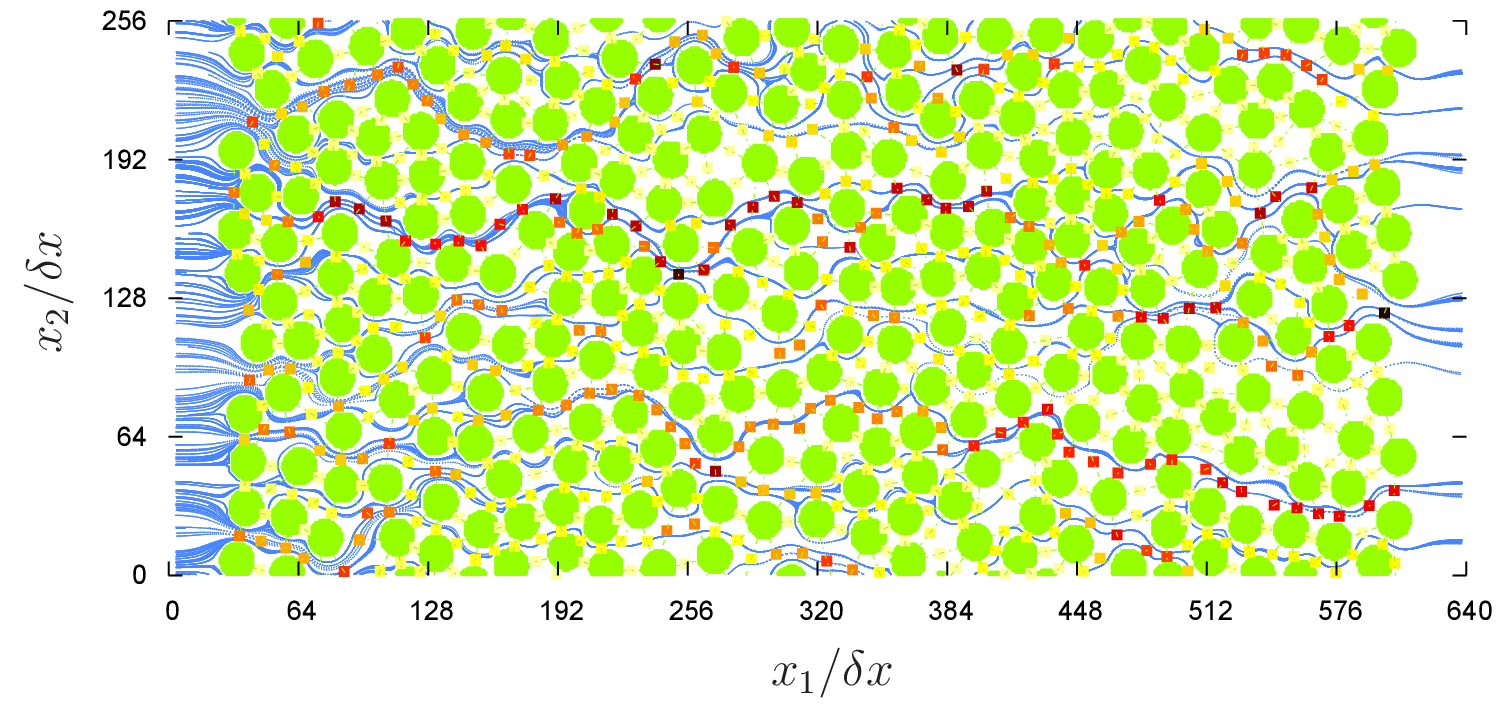}
  }
  \subfigure[$r_{\rm p}=2.6$]{ \label{fig:trj4}
    \includegraphics[width=0.48\textwidth]{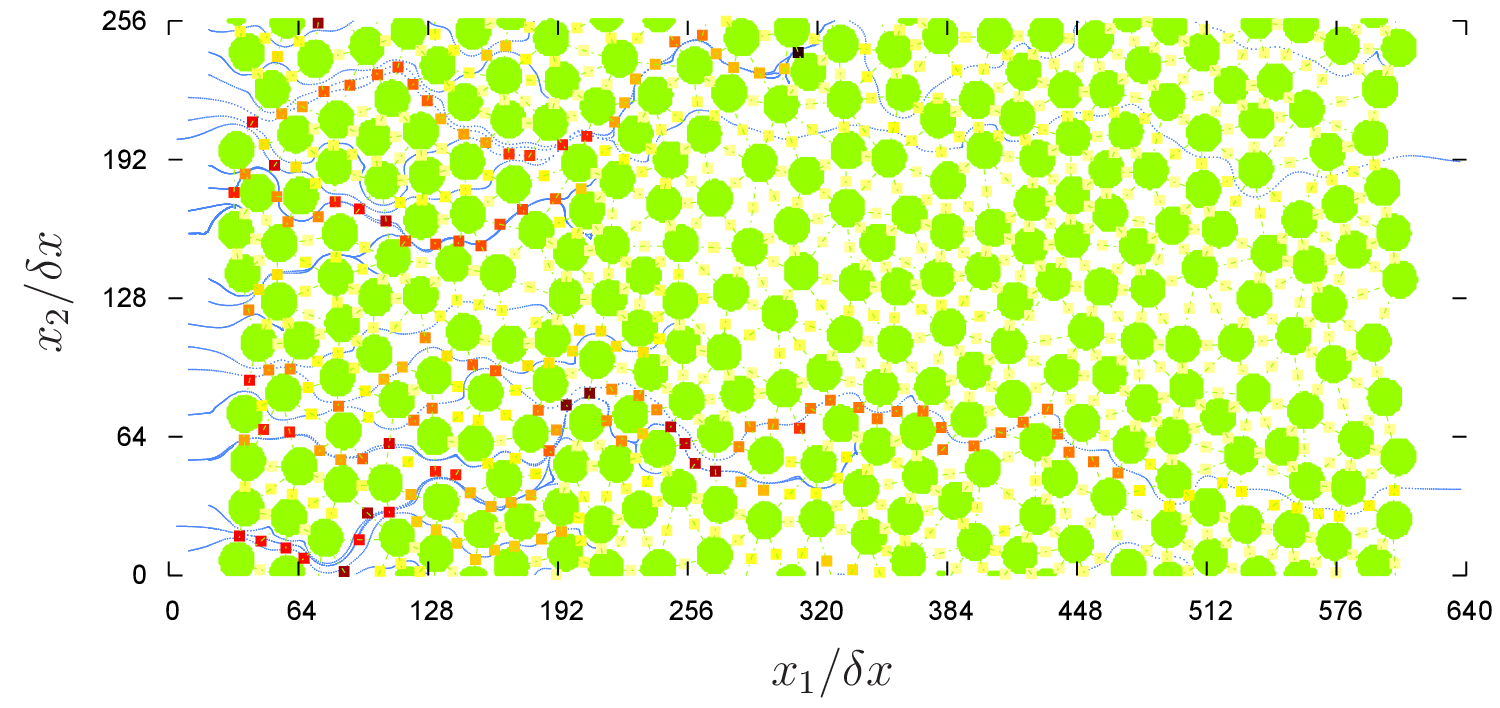}
  }
\caption{\label{fig:traj} {Time averaged steady state particle
    trajectories (blue lines) in a porous medium made by randomly
    placed cylinders (green circles). Between each pair of neighbor
    cylinders a colored square is located showing the probability of a
    particle to flow through. The radius of the particles
    $r_{\mathrm{p}}$ is varied between zero (tracers) and $2.6$.}
}
\end{figure}

\subsection{Particle suspensions}
Particle suspensions in porous media are relevant in many processes
such as fines migration~\cite{Leonardi2012a,Leonardi2012b}, sand liberation~\cite{Boutt2011}, catalysis, heap-leaching, filtration, fertilization, and
contaminant spreading in the subsoil. In these applications, relevant
questions involve the possibility of sealing of porous media,
segregation, or the formation of particle clusters and their influence
on the transport properties inside the porous structure. While for
some applications one might like to optimize a porous medium so as to
allow almost all particles to pass (e.g. reactors and
catalysts), other applications demand a perfect trapping of all
particles (e.g. filters).

Coupled lattice Boltzmann and discrete element/molecular dynamics algorithms are a
powerful tool to simulate such systems since they leverage the strengths of two numerical algorithms. As demonstrated in the previous
sections, the LBM is well suited to describe
(multiphase) fluid flow in complex geometries. Particle based methods, on the other hand, allow the description of interacting particles by solving Newton's equations of motion.  Here, we limit ourselves to
non-interacting point-wise particles (tracers) and massive spherical
particles which only interact through hydrodynamic and Hertz forces in
order to mimic hard spheres. The massive particles have the same mass
density as the fluid.  However, this is not a general restriction of
the algorithm.  Electrostatic, van der Waals, magnetic, or any other
kind of interactions can be used in the same way as in classical
molecular dynamics. We also ignore te effect of diffusion on tracer particles which could be taken care of by adding a diffusive term~\cite{Yang:2013b}.

\begin{figure}
  \subfigure[tracer]{\label{fig:hist1}
   \includegraphics[width=.48\linewidth]{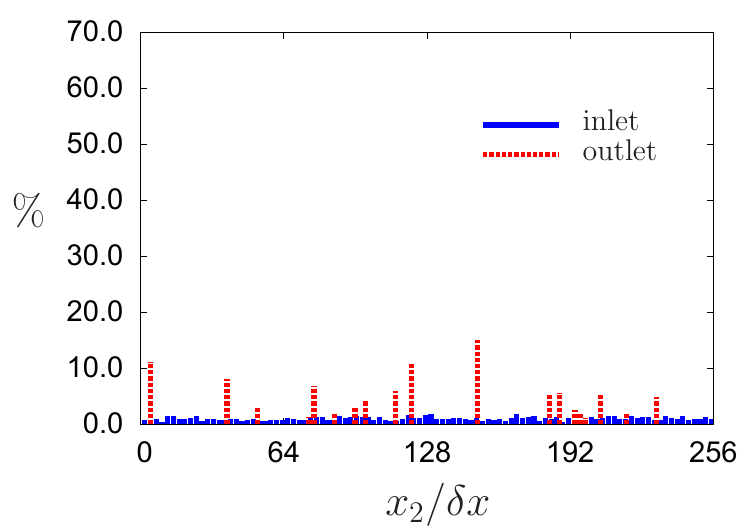}
  }
  \subfigure[$r_{\mathrm{p}}=1.6$]{ \label{fig:hist2}
   \includegraphics[width=.48\linewidth]{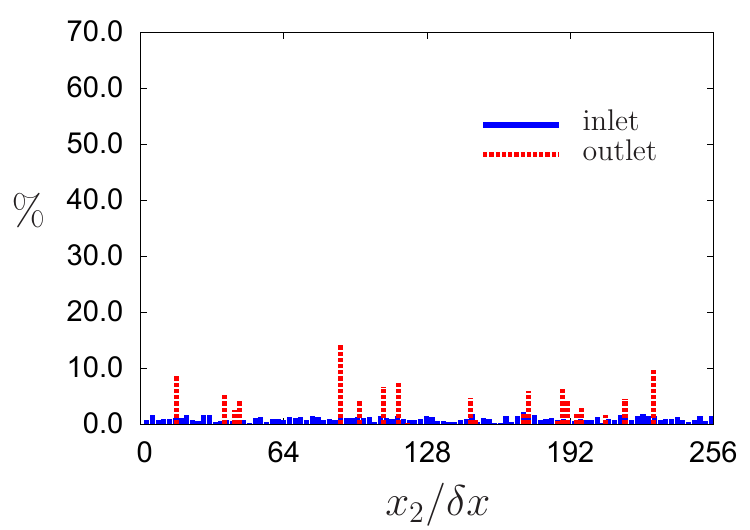}
  }

  \subfigure[$r_{\mathrm{p}}=2.1$]{ \label{fig:hist3}
    \includegraphics[width=.48\linewidth]{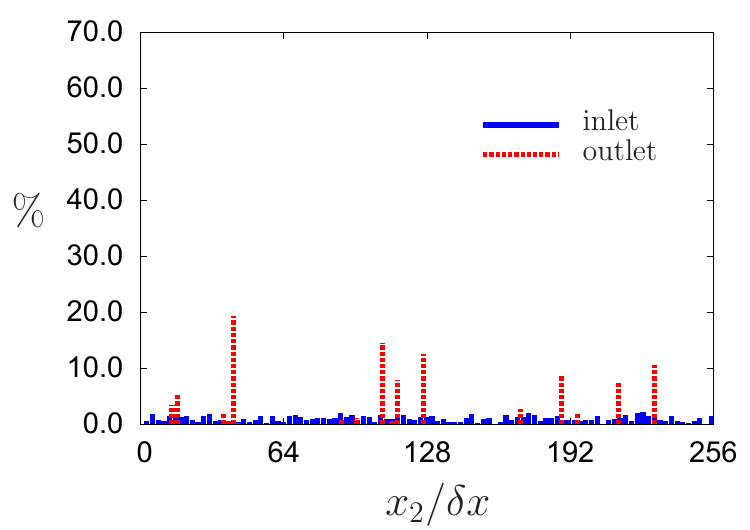}
  }
  \subfigure[$r_{\mathrm{p}}=2.6$]{ \label{fig:hist4}
    \includegraphics[width=.48\linewidth]{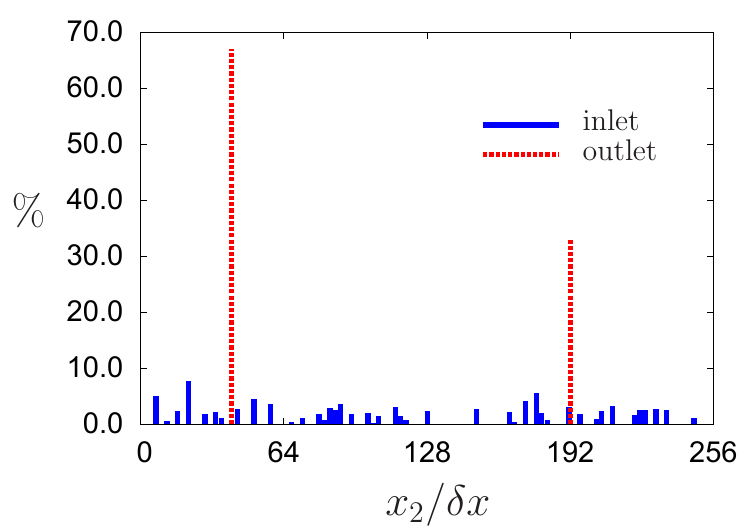}
  }
\caption{\label{fig:histInOut} {Histogram of the particle positions at
  the inlet and outlet plane of the model porous medium for tracers
  and massive particles of radius $r_{\mathrm{p}}=1.6$,
  $r_{\mathrm{p}}=2.1$, and $2.6$, respectively.}
}
\end{figure}
Our system of interest is again a pseudo-2D porous medium made of randomly
placed cylinders as shown in Fig.~\ref{fig:traj}. The system size is
$256\times640$ and the fluid is driven using pressure boundary conditions
in $x_1$ direction. Particles leaving the simulation domain at the outlet in
$x_1$ direction re-enter at the inlet, but at a randomly chosen $x_2$ position.
All other boundaries are periodic. These particles only interact with walls by
means of lubrication forces and a very short range repulsive force. When the
simulation has reached a steady state, we record the trajectories of 1000
tracer particles or 100 massive particles being transported by the flow.
{Figs.~\ref{fig:trj1} to \ref{fig:trj4} depict these trajectories for tracers
and particles with radius, $r_{\mathrm{p}}=1.6$, $2.1$, and $2.6$,
respectively.} It can be clearly seen that even for tracer particles preferable
paths exists. This is due to high local flow velocities which have their
origin in the particular arrangement of the cylinders. When increasing the
particle radius, the number of preferable paths reduces for several reasons.
First, particles with an extended size are only able to pass through pores
which are larger than the particle diameter. Second, particles might block
small pores rendering the area behind it practically inaccessible for all
further particles flowing in. This effectively leads to a dynamic
rearrangement of the flow field and based on that the preferable paths the
particles tend to follow will change as well.

\begin{figure}
\centerline{\includegraphics[width=0.6\textwidth]{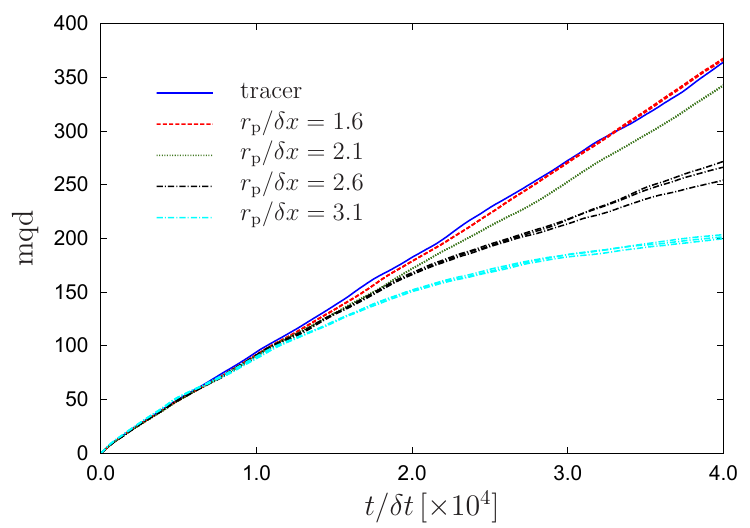}}
\caption{\label{fig:mqd} Mean square displacement of the particles
  inside the porous medium for different particle sizes.}
\end{figure}

This explanation is underlined by Figures \ref{fig:histInOut} and \ref{fig:mqd}. Fig.~\ref{fig:histInOut}
depicts histograms of $x_2$ positions where particles enter at the
inlet (randomly chosen) and where they leave the system at the
outlet. Interestingly, the tracers and the small particles with
$r_{\mathrm{p}}=1.6$ show a similar number of preferred outlet
positions. However, these are differently distributed because the
massive particles also influence the flow field -- even though they
are sufficiently small to pass almost all pore throats. The histogram for the largest particles with $r_{\mathrm{p}}=2.6$ shows that only three possible percolating paths seem to be accessible, while all others include pore throats which are smaller than the particle diameter.

By averaging over all particles $i$ the recorded trajectories can be
used to compute the mean square displacement
$\mathrm{mqd}(t)=\left<(r_i(t)-r_i(t=0))^2\right>_{i,t}$. $\mathrm{mqd}(t)$
is shown for the different particle species in Fig.~\ref{fig:mqd}.
It can be concluded that small massive particles can be transported
through the porous medium almost as efficiently as tracers -- even
though they follow different preferable paths as demonstrated by the
histograms in Fig.~\ref{fig:histInOut}. For larger particles, however,
the probability that some of them get stuck in small pores grows
leading to a substantial reduction of the mean square displacement.

\section{Conclusion}
\label{sec:summ}
In this article we provided a comprehensive overview and literature review on
lattice Boltzmann modelling of multiphase flow with a particular focus on
porous media applications. We introduced several algorithmic extensions of the
LBM to describe multiple fluid phases or solid phases
suspended in fluid. Their individual advantages and disadvantages were
discussed based on simple example cases.

To guide readers to choose appropriately among the different LBM formulations
for multiple fluid phases reviewed above, Table~\ref{tab1} gives a brief
summary of their capabilities which are examined through a series of
comparisons, including (i) the ability of modeling the interfacial tension,
which can be given directly in the model or should be obtained numerically
through the static bubble test based on the Laplace's law, (ii) the magnitude
of maximum spurious velocities in the static bubble test, (iii) dissolution
rate for small droplets/bubbles, (iv) the highest density ratio that can be
achieved, (v) the highest kinematic viscosity ratio that can be achieved, and
(vi) the computing cost. As can be seen from Table~\ref{tab1}, each lattice
Boltzmann multiphase model has its own advantages and limitations, and it is
not possible to state that one model is definitely preferred to another.
However, it will be beneficial to be aware of and carefully consider the
following points, especially when the LBM is chosen for
pore-scale simulation of multiphase flows in porous media:

(1) The stabilized diffuse-interface model can almost eliminate the spurious
velocities to round-off error, free-energy and color gradient models produce larger spurious velocities and the inter-particle potential model has largest spurious velocities;

(2) Small droplets/bubbles are expected to dissolve for stabilized diffuse-interface model, free-energy model and mean-field theory model, and the dissolution rate is fastest for the mean-field model;

(3) The stabilized diffuse-interface model is most suitable to simulate  flow
problems with high density ratio, while the color gradient model is most suitable to simulate flows with moderate/high viscosity ratio;

(4) The free-energy model of potential form can produce smaller spurious velocities than its stress/pressure form, thus leading to the correct equilibrium contact angles when the binary fluids have different viscosities.

\begin{table}[!htp]
 \caption{A summary of the capabilities of several lattice Boltzmann multiphase models.}
\label{tab1}
\begin{tabular}{c|ccc}
\hline\hline &  & Models   &\\
\cline{2-4}
 &Inter-particle potential & Color&  Free-energy \\
\hline
Interfacial tension & static bubble test & can be given & can be given  \\
& is required & directly & directly  \\
Spurious currents & large & medium & small \\
Dissolution for tiny & small & very small & large  \\
bubbles/droplets &  &  &  \\
Density ratio & 1000$^\dag$ & 1000$^{*}$ & 1  \\
Kinematic viscosity & 1000$^\ddag$ & 1000 & up to 8  \\
ratio &  &  &   \\
Computing cost & `average' & `average' & `average'  \\
\hline\hline &  & Models   &\\
\cline{2-4}
 &Mean-field theory & Stabilized diffuse-interface &\\
\hline
Interfacial tension & static bubble & can be given &\\
 & test is required & directly & \\
Spurious currents &  medium & very small & \\
Dissolution for tiny &  very large & medium &\\
bubbles/droplets &  &  &   \\
Density ratio &  up to around 15 & 1000 &\\
Kinematic viscosity & unknown & unknown &\\
ratio &  &  &  \\
Computing cost & greater & large$^{**}$ &\\
\hline
\end{tabular}
$^\dag$ Achieved in static bubble test with both SCMP~\cite{Yuan2006} and MCMP~\cite{Bao2013} models using equations of state different from the original Shan-Chen model. \\
$^\ddag$ Achieved in static bubble test and two-phase cocurrent flow between two parallel plates with the MCMP model, using higher-order isotropy in the fluid-fluid interfacial terms,
explicit forcing scheme, and multiple relaxation times~\cite{Porter2012}. \\
$^{*}$ Achieved in static bubble test using the color gradient model presented in ~\cite{Liu2012a}.\\
$^{**}$ The normal direction at each boundary node should be identified, and high-order approximations to derivatives are needed.
\end{table}

\begin{acknowledgements}
We thank Vahid Joekar Niasar and Cor van Kruijsdijk for the organisation of the workshop on ``(sub) pore-scale modeling of multiphase flow and transport in porous media'' which took place in January 2013. C. Leonardi and J. Williams acknowledge the support of Schlumberger Doll Research. A. Narv\'aez, S. Schmieschek and J. Harting acknowledge financial support from
NWO/STW (Vidi grant 10787 of J.~Harting) and FOM/Shell IPP (09iPOG14 -
``Detection and guidance of nanoparticles for enhanced oil recovery''). We
thank the J\"ulich Supercomputing Centre, Sara Amsterdam and HLRS Stuttgart for
computing resources. 
H. Liu, Q. Kang, and A. Valocchi acknowledge the support from the LDRD Program
and Institutional Computing Program of the Los Alamos National Laboratory.  H.
Liu and A. Valocchi gratefully acknowledge additonal support of the
International Institute for Carbon Neutral Energy Research (WPI-I2CNER),
sponsored by the Japanese Ministry of Education, Culture, Sports, Science and
Technology. H. Liu would like to thank the financial support from the Thousand Youth Talents Program for Distinguished Young Scholars, China.
\end{acknowledgements}

\end{document}